\pdfoutput=1
\documentclass[aoas]{imsart}

\RequirePackage{amsthm,amsmath,amsfonts,amssymb}
\RequirePackage[authoryear]{natbib}
\RequirePackage{graphicx}
\RequirePackage{booktabs}
\RequirePackage{array}
\RequirePackage{float}
\RequirePackage{algorithm}
\RequirePackage{algpseudocode}
\RequirePackage[section]{placeins}

\startlocaldefs
\theoremstyle{plain}
\newtheorem{theorem}{Theorem}
\newtheorem{proposition}{Proposition}

\theoremstyle{definition}

\newcolumntype{L}[1]{>{\raggedright\arraybackslash}p{#1}}

\newcommand{\BatEtaRef}{0.005}
\newcommand{\BatNumTrips}{7}
\newcommand{\BatNumLocations}{5,271}
\newcommand{\BatNumSteps}{5,264}
\newcommand{\BatNumBuzzes}{269}
\newcommand{\BatNumBuzzLocations}{255}
\newcommand{\BatMeanDurationMin}{198.4}

\newcommand{\BatHMMLogLik}{-29,631.3}
\newcommand{\BatHMMAIC}{59,280.6}
\newcommand{\BatForagingStepMean}{51.8}
\newcommand{\BatCommutingStepMean}{84.8}
\newcommand{\BatForagingKappa}{0.001}
\newcommand{\BatCommutingKappa}{9.72}

\newcommand{\BatTubeRobustForagingPctEta}{28.2}
\newcommand{\BatTubeAmbiguousPctEta}{17.9}
\newcommand{\BatTubeRobustCommutingPctEta}{53.9}

\newcommand{\BatBuzzEnrichmentRobustForaging}{2.25}
\newcommand{\BatBuzzEnrichmentRobustForagingCI}{(1.73, 2.85)}
\newcommand{\BatBuzzEnrichmentAmbiguous}{1.23}
\newcommand{\BatBuzzEnrichmentAmbiguousCI}{(0.84, 1.56)}
\newcommand{\BatBuzzEnrichmentRobustCommuting}{0.27}
\newcommand{\BatBuzzEnrichmentRobustCommutingCI}{(0.16, 0.44)}
\newcommand{\BatBuzzRateRobustForaging}{0.115}

\newcommand{\BatBuzzRateAmbiguous}{0.063}

\newcommand{\BatBuzzRateRobustCommuting}{0.014}

\newcommand{\BatRateRatioAmbiguousVsRobustForaging}{0.58}
\newcommand{\BatRateRatioAmbiguousVsRobustForagingCI}{(0.37, 0.90)}
\newcommand{\BatRateRatioCommutingVsRobustForaging}{0.13}
\newcommand{\BatRateRatioCommutingVsRobustForagingCI}{(0.08, 0.22)}
\newcommand{\BatRateModelType}{quasi-Poisson}

\newcommand{\BatViterbiForagingEnrichment}{2.25}
\newcommand{\BatViterbiCommutingEnrichment}{0.34}
\newcommand{\BatMarginalForagingEnrichment}{2.33}

\newcommand{\BatMarginalCommutingEnrichment}{0.27}
\newcommand{\BatViterbiForagingRobustRate}{0.115}
\newcommand{\BatViterbiForagingAmbiguousRate}{0.116}
\newcommand{\BatViterbiCommutingAmbiguousRate}{0.033}
\newcommand{\BatViterbiCommutingRobustRate}{0.014}
\newcommand{\BatEtaLowRobustForagingEnrichment}{2.33}
\newcommand{\BatEtaHighRobustForagingEnrichment}{2.65}
\newcommand{\BatEtaLowAmbiguousPct}{8.6}
\newcommand{\BatEtaHighAmbiguousPct}{79.2}
\newcommand{\BatEtaLowRobustCommutingEnrichment}{0.27}
\newcommand{\BatEtaHighRobustCommutingEnrichment}{0.07}

\newcommand{\BatRepresentativeTrip}{Viv5\_flight1\_597}
\endlocaldefs

\begin{document}

\begin{frontmatter}
\title{Tropical Viterbi Tubes for Decoding Uncertainty in Hidden Markov Models}
\runtitle{Tropical Viterbi Tubes}

\begin{aug}
\author[A]{\fnms{Aur\'elien}~\snm{Nicosia}\ead[label=e1]{nicosia.aurelien@gmail.com}}
\address[A]{Universit\'e Laval, Qu\'ebec City, QC G1V 0A6, Canada\printead[presep={,\ }]{e1}}
\end{aug}

\begin{abstract}
Hidden Markov models are widely used to infer latent state sequences from
sequential data, and Viterbi decoding returns a single most likely hidden
trajectory. In applications where decoded states have scientific meaning, this
trajectory is a point estimate of a high-dimensional latent object. When
near-optimal complete paths coexist, a single maximizer can obscure pathwise
uncertainty. Conditional on a fitted HMM, we introduce the \emph{tropical
Viterbi tube} as a pathwise uncertainty set around Viterbi decoding: the set of
hidden trajectories whose complete-data log-score lies within tolerance of the
Viterbi optimum. Its state, transition, and change-status projections summarize
which local features are compatible with near-optimal complete paths.

Conditional on a fitted HMM, the tropical Viterbi tube is a posterior superlevel
set on the complete hidden-path space, with tolerance interpreted as a log
posterior-odds loss relative to a Viterbi path. When calibrated to attain a
target posterior mass, the tube becomes a posterior superlevel, or
HPD-threshold, credible region for the complete latent path, with possible
conservatism due to discreteness and boundary ties; its projections yield
conservative simultaneous credible bands. We establish monotonicity,
step-function properties of projected summaries, and deterministic stability
guarantees showing how tube width controls changes in Viterbi decoding under
bounded score perturbations.

Projected tubes are computed exactly by max-plus forward--backward recursions.
For dense transition matrices, state and transition entrance tolerances are
obtained in \(O(TK^2)\) time, while the full posterior mass of the tube is
treated separately because it requires a global path-score constraint.
The \(O(TK^2)\) computation is for projected tubes and entrance tolerances;
posterior tube mass and HPD mass calibration are separate pathwise calculations
and are approximated by FFBS in the numerical work. In a
public bat-tracking application with acoustic data on feeding buzzes, which
provide external evidence of prey-capture attempts rather than continuous state
annotations, robust foraging tube segments are enriched for observed feeding
buzzes and robust commuting segments are depleted. At \(\eta=\BatEtaRef\),
robust foraging enrichment is
\BatBuzzEnrichmentRobustForaging{} with 95\% bootstrap interval
\BatBuzzEnrichmentRobustForagingCI, whereas robust commuting enrichment is
\BatBuzzEnrichmentRobustCommuting{} with interval
\BatBuzzEnrichmentRobustCommutingCI. In the bat application, the main added value is not improved point classification of foraging, but separation of stable commuting periods from ambiguous Viterbi-commuting intervals where near-optimal complete paths remain compatible with alternative behaviour.
\end{abstract}

\begin{keyword}
\kwd{Hidden Markov models}
\kwd{Viterbi decoding}
\kwd{tropical geometry}
\kwd{max-plus algebra}
\kwd{decoding uncertainty}
\kwd{posterior path uncertainty}
\end{keyword}

\end{frontmatter}

\begin{center}
\textit{Preprint version. Submitted to The Annals of Applied Statistics. June 15, 2026.}
\end{center}

\section{Introduction}

Hidden Markov models (HMMs) are among the standard statistical tools for
sequential data in which the observed process is driven by an unobserved
finite-state process. They are used in speech and signal processing, biological
sequence analysis, ecological time series, animal movement studies, finance, and
medical monitoring \citep{rabiner1989tutorial,durbin1998biological,
cappe2005inference,zucchini2016hidden,langrock2012flexible,
mcclintock2020uncovering}. In many of these applications, the hidden states are
not merely auxiliary variables introduced for computational convenience. They
carry substantive scientific meaning: behavioural modes of an animal,
functional genomic regions, market regimes, or latent phases of a physiological
process. After an HMM has been fitted, a central inferential task is therefore
decoding, namely assigning plausible hidden-state trajectories to the observed
sequence.

The most widely used decoding method is the Viterbi algorithm
\citep{viterbi1967error,rabiner1989tutorial}. In an HMM, the Viterbi decoder
returns a complete hidden-state path with maximum posterior probability,
equivalently a path with maximum complete-data score under the fitted model
\citep{cappe2005inference,zucchini2016hidden}. This global nature is one of its
main advantages. Unlike a collection of pointwise decisions, the Viterbi path is
a coherent trajectory that respects the transition structure of the model. Its
computational cost is also modest, making it attractive for long time series and
large numbers of fitted models. For these reasons, the Viterbi path is routinely
used as the decoded sequence in applied analyses.

However, the Viterbi path is a point estimate of a high-dimensional latent
object. This creates a statistical difficulty that is often hidden by the
algorithm's apparent definitiveness. When several complete trajectories have
nearly equal posterior support, reporting a single maximizer may give an overly
certain description of the latent process. Small changes in estimated
parameters, in observation likelihoods, or in numerical implementation may alter
the selected path even when the competing explanations are scientifically
similar. Conversely, two paths may differ at a few important transition times
while having almost indistinguishable complete-data scores. In such situations,
the inferential question is not only which path is optimal, but also which
alternative paths are nearly optimal, where they differ from the Viterbi path,
and whether the scientific conclusions drawn from decoding are stable. We refer
to this problem as \emph{path uncertainty}: uncertainty about the complete
latent trajectory as a structured object.

Existing summaries of HMM uncertainty only partially address this problem. The
forward--backward algorithm provides posterior marginal probabilities for each
hidden state at each time point \citep{rabiner1989tutorial,cappe2005inference}.
These marginals are useful for local diagnostics and for posterior decoding, but
they do not directly describe uncertainty over complete paths. A sequence of
locally plausible states need not itself correspond to a high-probability
trajectory under the HMM transition structure. Entropy-based summaries provide
another important view of uncertainty \citep{hernando2005efficient}. Local
entropy summarizes dispersion in the marginal state distribution, whereas
state-sequence entropy gives a scalar measure of uncertainty in the hidden path.
These quantities are informative, but they do not identify which complete
trajectories are plausible competitors to the Viterbi path, nor do they yield a
simultaneous band for states, transitions, or change points. More generally, the
distinction between Viterbi decoding, posterior decoding, and risk-based path
inference is well recognized \citep{lember2014bridging}, but there remains a
need for a computationally tractable object that represents uncertainty at the
level of complete trajectories.

Related algorithmic work has considered exploration of the hidden
state-sequence space through generalized or list Viterbi algorithms, \(k\)-best
paths, and sampling-based methods
\citep{seshadri1994list,guedon2007exploring,brown2010decoding}. These methods
are useful for enumerating or sampling alternative explanations and for studying
structural differences between high-probability paths. They do not, by
themselves, define a posterior probability level set around the Viterbi optimum,
nor do they provide entrance tolerances and simultaneous projected state,
transition, and change-status bands. The construction developed below is
therefore complementary to list-based and sampling-based exploration of the
state-sequence space. A \(k\)-best algorithm could be stopped once path scores
fall below \(\psi^\star-\varepsilon\), but this would enumerate paths and would
not by itself provide the \(O(TK^2)\) projected entrance profiles used here.

\begin{table}[t]
\centering
\caption{Positioning of tropical Viterbi tubes relative to common HMM decoding
and uncertainty summaries.}
\label{tab:method-positioning}
\scriptsize
\setlength{\tabcolsep}{2pt}
\renewcommand{\arraystretch}{1.15}
\begin{tabular}{@{}L{0.125\linewidth}L{0.095\linewidth}L{0.115\linewidth}L{0.110\linewidth}L{0.095\linewidth}L{0.130\linewidth}L{0.115\linewidth}L{0.135\linewidth}@{}}
\toprule
Feature & Viterbi & Marginals & \shortstack[l]{Posterior\\decoding} & Entropy & \(k\)-best/list & \shortstack[l]{FFBS\\paths} & \shortstack[l]{Tropical\\tube}\\
\midrule
Object returned & one MAP path & \(P(S_t=k\mid Y)\) & pointwise MAP states & scalar or local uncertainty & ranked high-score paths & posterior path samples & score-threshold path set and exact projections\\
Pathwise? & yes & no & no & no & yes & yes & yes\\
Local posterior? & no & yes & yes & yes & no & yes by MC & no\\
Globally coherent? & yes & no & not necessarily & no & yes & yes & yes\\
Exact projected bands? & no & no & no & no & by post-processing only & MC only & yes, conservative\\
Entrance tolerances? & no & no & no & no & not directly & not directly & yes\\
Posterior mass? & no & local only & no & no & partial lower bound & yes by MC & separate calculation\\
\bottomrule
\end{tabular}
\renewcommand{\arraystretch}{1}
\end{table}

The proposed contribution is therefore not merely the existence of high-scoring
alternatives to the Viterbi path. Such alternatives can be enumerated or
sampled by existing methods. The contribution is the set-valued pathwise object
together with exact projected membership, entrance tolerances, gap diagnostics,
and simultaneous projected-band interpretation. These quantities summarize
which local scientific statements are stable across all globally near-optimal
complete paths, without enumerating the tube.

This paper introduces such an object: the \emph{tropical Viterbi tube}. For a
given tolerance, the tube is the set of all hidden-state trajectories whose
complete-data score lies within that tolerance of the Viterbi optimum. At zero
tolerance, it contains all Viterbi-optimal paths. At positive tolerance, it
contains paths that are not necessarily optimal but are close to optimal in the
same global sense used by the Viterbi algorithm. The term ``tropical'' reflects
the max-plus algebraic structure of the Viterbi recursion
\citep{baccelli1992synchronization,pachter2004tropical,
maclagan2015introduction,theodosis2018analysis,maragos2021tropical}. The
statistical idea, however, is simple: replace a single decoded path by a set of
globally near-optimal paths.

The tube differs from local uncertainty sets in an essential way. A state is
included in the projected tube at a given time only if there exists an entire
hidden trajectory, coherent before and after that time, whose score remains
close to the Viterbi score. Similarly, a transition is included only if it
belongs to at least one globally near-optimal trajectory. Thus the proposed
object preserves the pathwise structure of the HMM. It can show, for example,
that a state with moderate posterior marginal probability is not compatible with
any near-optimal complete path, or that two alternative change-point locations
are both supported by globally coherent explanations. This distinction is
important in applications where the decoded path is interpreted scientifically.

The contributions of this paper are as follows. First, conditional on a fitted
HMM, we define the tropical Viterbi tube as a pathwise set-valued object around
Viterbi decoding. Second, we show that the tube is a posterior superlevel set
on the complete hidden-path space and that, after posterior mass calibration,
it becomes a posterior superlevel or HPD-threshold credible region with the
usual discreteness and boundary-tie conservatism. Third, we give exact
projected state, transition, and change-status membership, entrance tolerances,
and gap diagnostics. The main exact computational contribution is not
enumeration of the full tube, but computation of its state, transition and
change-status projections and entrance tolerances. Fourth, we distinguish
average set-valued calibration, simultaneous projected-band calibration, and
HPD path-mass calibration, emphasizing that projected simultaneous bands are
conservative summaries of a pathwise object. Fifth, we establish deterministic
stability guarantees under bounded perturbations of the complete-data score.
Finally, the simulations and bat application show how the tube summarizes
pathwise stability, rather than replacing Viterbi, posterior marginals, or
entropy as a point classifier.

The practical consequence is that decoding can be presented as a stability
analysis rather than as a single classification. In an animal movement
study, for example, a fitted HMM may classify locations into resting, foraging,
and travelling states \citep{langrock2012flexible,mcclintock2020uncovering}. A
Viterbi path gives one behavioural history. The tropical Viterbi tube instead
identifies the set of behavioural histories that are nearly as well supported by
the fitted model. Time intervals where the tube projection is a singleton
correspond to stable decoded behaviour. Intervals where the projection widens
indicate genuine pathwise ambiguity. Transition projections similarly
distinguish robust changes of behaviour from uncertain change-point locations.
These summaries are directly useful for interpretation, for assessing robustness
of ecological conclusions, and for decisions based on decoded states. The
application in Section~\ref{sec:application} uses GPS and acoustic recordings
from Mexican fish-eating bats to show how projected tubes distinguish stable
foraging segments from ambiguous transitions in a two-state movement HMM.

The remainder of the paper moves from notation to interpretation, computation,
and empirical use. Section~\ref{sec:hmm-score-geometry} sets up HMM decoding,
posterior path probabilities, Viterbi score gaps, and the max-plus recursion.
Section~\ref{sec:tropical-viterbi-tubes} defines tropical Viterbi tubes and
their projected uncertainty summaries. Section~\ref{sec:theory} gives the
statistical interpretation and guarantees, including posterior superlevel sets,
HPD path calibration, simultaneous projected bands, and perturbation stability.
Section~\ref{sec:computation} gives the exact computation of projected tubes
and discusses the separate problem of posterior tube mass. Section~\ref{sec:simulation}
reports the simulation study. Section~\ref{sec:application} reports the bat
movement application, and Section~\ref{sec:discussion} discusses limitations
and extensions. Additional geometric details are given in the Supplement.

\section{HMM Decoding and Viterbi Score Geometry}
\label{sec:hmm-score-geometry}

This section fixes notation and formulates Viterbi decoding as a max-plus
optimization problem. The point of this formulation is not merely algebraic. It
identifies the complete hidden-state trajectory as the statistical object being
estimated, clarifies the posterior meaning of Viterbi score gaps, and provides
the computational structure used later to construct projected tropical Viterbi
tubes.

\subsection{Fitted HMM and complete-data path scores}
\label{subsec:fitted-hmm-score}

Let \(S_t \in \mathcal S=\{1,\ldots,K\}\), \(t=1,\ldots,T\), denote the latent
state process, and let \(Y_{1:T}=y_{1:T}\) denote the observed sequence.
Throughout the theoretical development we condition on a fitted HMM, with
parameter value denoted implicitly by \(\widehat\theta\). Thus all probabilities
and densities below are conditional on the fitted model. This conditional
viewpoint is standard in HMM decoding: parameter estimation and model
uncertainty are important, but the decoding problem first asks what can be
inferred about \(S_{1:T}\) given the fitted model and the observed sequence
\citep{rabiner1989tutorial,cappe2005inference,zucchini2016hidden}.

We allow time-inhomogeneous transition probabilities in order to include
covariate-dependent or nonstationary HMMs. The time-homogeneous case is obtained
by setting the transition matrix independent of \(t\). Write
\[
  \pi_k = \Pr(S_1=k), \qquad
  \gamma_t(i,j)=\Pr(S_t=j \mid S_{t-1}=i), \qquad t=2,\ldots,T,
\]
and let \(f_{t,j}(y_t)\) be the emission probability mass or density of the
observation at time \(t\) in state \(j\). We work on the log scale and define
\[
  p_k=\log \pi_k, \qquad
  a_t(i,j)=\log \gamma_t(i,j), \qquad
  e_t(j)=\log f_{t,j}(y_t),
\]
with the convention \(\log 0=-\infty\). A path with any impossible initial
state, transition, or emission therefore receives score \(-\infty\).

For a hidden path \(s_{1:T}\in\mathcal S^T\), define the complete-data log-score
\begin{equation}
\label{eq:complete-data-score}
  \psi(s_{1:T};y_{1:T})
  =
  p_{s_1}+e_1(s_1)
  +
  \sum_{t=2}^T
  \left\{
    a_t(s_{t-1},s_t)+e_t(s_t)
  \right\}.
\end{equation}
When the emissions are discrete, this is the log joint probability of
\((S_{1:T},Y_{1:T})=(s_{1:T},y_{1:T})\). When the emissions are continuous, it
is the corresponding log joint density in \(y_{1:T}\) and probability mass in
the hidden path. In either case, the posterior distribution over the finite path
space is
\begin{equation}
\label{eq:path-posterior}
  \Pi_y(s_{1:T})
  :=
  \Pr(S_{1:T}=s_{1:T}\mid Y_{1:T}=y_{1:T})
  =
  \exp\{\psi(s_{1:T};y_{1:T})-\ell(y_{1:T})\},
\end{equation}
where
\begin{equation}
\label{eq:observed-loglik}
  \ell(y_{1:T})
  =
  \log
  \sum_{r_{1:T}\in\mathcal S^T}
  \exp\{\psi(r_{1:T};y_{1:T})\}
\end{equation}
is the observed-data log-likelihood under the fitted HMM. Terms with score
\(-\infty\) contribute zero to the sum. We assume throughout that at least one
path has finite score.

In what follows, statements involving posterior probabilities, credible
regions, coverage under the fitted law, or decoding uncertainty are all
conditional statements under this fitted model unless explicitly stated
otherwise.

\subsection{The Viterbi path as a posterior mode}
\label{subsec:viterbi-mode}

The Viterbi score is
\begin{equation}
\label{eq:viterbi-score}
  \psi^\star(y_{1:T})
  =
  \max_{s_{1:T}\in\mathcal S^T}
  \psi(s_{1:T};y_{1:T}).
\end{equation}
The corresponding set of Viterbi-optimal paths is
\begin{equation}
\label{eq:viterbi-set}
  \mathcal V(y_{1:T})
  =
  \left\{
    s_{1:T}\in\mathcal S^T:
    \psi(s_{1:T};y_{1:T})=\psi^\star(y_{1:T})
  \right\}.
\end{equation}
Any element \(\widehat s_{1:T}\in\mathcal V(y_{1:T})\) is a Viterbi path.
Equivalently, since \(\ell(y_{1:T})\) does not depend on \(s_{1:T}\),
\[
  \mathcal V(y_{1:T})
  =
  \arg\max_{s_{1:T}\in\mathcal S^T}
  \Pi_y(s_{1:T}).
\]
Thus Viterbi decoding returns a posterior mode on the complete hidden-path space
\citep{viterbi1967error,forney1973viterbi,rabiner1989tutorial}.

This observation gives an immediate statistical interpretation to score
differences. For any \(s^\star_{1:T}\in\mathcal V(y_{1:T})\) and any path
\(s_{1:T}\),
\begin{equation}
\label{eq:deficit-posterior-odds}
  \log
  \frac{\Pi_y(s^\star_{1:T})}{\Pi_y(s_{1:T})}
  =
  \psi^\star(y_{1:T})-\psi(s_{1:T};y_{1:T}).
\end{equation}
Consequently, a path whose score is within \(\varepsilon\) of the Viterbi score
has posterior probability at least \(e^{-\varepsilon}\) times the posterior
probability of a Viterbi path. The tolerance scale used later in the tropical
Viterbi tube is therefore a log-posterior-odds scale, not an arbitrary numerical
threshold.

It is important to distinguish the Viterbi set from the single path returned by
an implementation. If the maximum in \eqref{eq:viterbi-score} is attained by
more than one path, a backtracking convention may select one of them
arbitrarily. Such tie-breaking is harmless for obtaining one maximizer, but it
is statistically relevant when decoding is used as a scientific summary. The
tube construction developed below treats ties and near-ties as part of the
inferential object rather than as numerical nuisances.

\subsection{Dynamic programming recursion}
\label{subsec:viterbi-dp}

The Viterbi algorithm computes \(\psi^\star(y_{1:T})\) without enumerating the
\(K^T\) possible hidden paths. For a partial path \(s_{1:t}\), define the
partial score
\[
  \psi_t(s_{1:t};y_{1:t})
  =
  p_{s_1}+e_1(s_1)
  +
  \sum_{u=2}^t
  \{a_u(s_{u-1},s_u)+e_u(s_u)\}.
\]
Let
\begin{equation}
\label{eq:viterbi-forward-score}
  F_t(j)
  =
  \max_{s_{1:t-1}\in\mathcal S^{t-1}}
  \psi_t(s_{1:t-1},s_t=j;y_{1:t})
\end{equation}
be the best partial score among paths ending in state \(j\) at time \(t\).
Then
\begin{equation}
\label{eq:viterbi-init}
  F_1(j)=p_j+e_1(j),
  \qquad j=1,\ldots,K,
\end{equation}
and, for \(t=2,\ldots,T\),
\begin{equation}
\label{eq:viterbi-recursion}
  F_t(j)
  =
  e_t(j)
  +
  \max_{i\in\mathcal S}
  \left\{
    F_{t-1}(i)+a_t(i,j)
  \right\}.
\end{equation}
Finally,
\begin{equation}
\label{eq:viterbi-final}
  \psi^\star(y_{1:T})=\max_{j\in\mathcal S} F_T(j).
\end{equation}
A Viterbi path can be recovered by storing predecessor sets
\[
  \mathcal B_t(j)
  =
  \arg\max_{i\in\mathcal S}
  \left\{
    F_{t-1}(i)+a_t(i,j)
  \right\},
  \qquad t=2,\ldots,T.
\]
When \(\mathcal B_t(j)\) has more than one element, there are multiple optimal
prefixes leading to state \(j\). Thus nonuniqueness can arise locally in the
recursion even before considering the full path.

For dense transition matrices, the recursion costs \(O(TK^2)\) operations. If
the transition graph is sparse and \(m_t\) transitions are allowed from time
\(t-1\) to time \(t\), the corresponding cost is \(O(\sum_{t=2}^T m_t)\), up to
the linear emission terms. This is the same basic computational scaling
exploited later for exact projected tube calculations.

\subsection{Max-plus form}
\label{subsec:maxplus-form}

The recursion \eqref{eq:viterbi-recursion} is a dynamic program over the
max-plus semiring. Let
\[
  u\oplus v=\max\{u,v\},
  \qquad
  u\otimes v=u+v,
\]
on \(\mathbb R\cup\{-\infty\}\), with additive identity \(-\infty\) and
multiplicative identity \(0\). Then \eqref{eq:viterbi-recursion} can be written
as
\begin{equation}
\label{eq:maxplus-recursion}
  F_t(j)
  =
  e_t(j)
  \otimes
  \bigoplus_{i\in\mathcal S}
  \left\{
    F_{t-1}(i)\otimes a_t(i,j)
  \right\}.
\end{equation}
This is the log-domain version of the max-product, or Viterbi, recursion. It is
parallel to the ordinary forward recursion, which uses summation over paths
rather than maximization. More generally, both recursions are instances of
distributive dynamic programming on factored models
\citep{aji2000generalized,kschischang2001factor}. The max-plus formulation is
also standard in idempotent algebra and discrete-event systems
\citep{baccelli1992synchronization}.

The distinction between sum-product and max-product is central for what
follows. The observed likelihood \(\ell(y_{1:T})\) in
\eqref{eq:observed-loglik} aggregates posterior mass over all paths. The
Viterbi score \(\psi^\star(y_{1:T})\) selects only the largest pathwise
contribution. Posterior marginal probabilities and entropies summarize the
sum-product posterior distribution locally, whereas the tropical Viterbi tube
will summarize the neighbourhood of the max-product optimum globally.

\subsection{Piecewise-linear structure of decoding}
\label{subsec:piecewise-linear-decoding}

For a fixed path \(s_{1:T}\), the score \(\psi(s_{1:T};y_{1:T})\) is affine in
the log-initial, log-transition, and log-emission scores appearing in
\eqref{eq:complete-data-score}. Therefore the Viterbi score is a finite maximum
of affine functions:
\[
  \psi^\star(y_{1:T})
  =
  \max_{s_{1:T}\in\mathcal S^T}
  \psi(s_{1:T};y_{1:T}).
\]
Consequently, the fitted score space is divided into regions on which a given
path, or set of tied paths, is optimal. Boundaries between such regions occur
when two or more complete paths have equal score. This is the piecewise-linear
geometry underlying the term ``tropical'' in this paper
\citep{pachter2004tropical,maclagan2015introduction,theodosis2018analysis}.

The present paper uses this geometry in a statistical way. We do not seek only
the region in which a single path is optimal. Instead, we study the collection
of paths whose scores lie close to the optimum, and we project that collection
onto time-indexed states, transitions, and change indicators. The max-plus
representation above is therefore the computational and conceptual bridge
between ordinary Viterbi decoding and the pathwise uncertainty object introduced
in the next section.

\section{Tropical Viterbi Tubes and Projected Uncertainty Summaries}
\label{sec:tropical-viterbi-tubes}

The usual output of Viterbi decoding is a single path, hence a point estimate
of the hidden trajectory. The tropical Viterbi tube augments that point estimate
with a pathwise near-optimality set.

For any path \(s_{1:T}\), define the Viterbi score deficit
\[
\Delta(s_{1:T};y_{1:T})
=
\psi^\star(y_{1:T})-\psi(s_{1:T};y_{1:T}).
\]
By \eqref{eq:deficit-posterior-odds}, this deficit is the log posterior odds
against \(s_{1:T}\) relative to any Viterbi-optimal path.
For a tolerance \(\varepsilon\geq 0\), define the tropical Viterbi tube by
\[
\mathcal T_\varepsilon(y_{1:T})
=
\left\{
s_{1:T}\in\mathcal S^T:
\Delta(s_{1:T};y_{1:T})\leq \varepsilon
\right\}.
\]
Equivalently,
\[
\mathcal T_\varepsilon(y_{1:T})
=
\left\{
s_{1:T}\in\mathcal S^T:
\psi(s_{1:T};y_{1:T})
\geq
\psi^\star(y_{1:T})-\varepsilon
\right\}.
\]
At \(\varepsilon=0\), the tube is the set of all Viterbi-optimal paths. For
\(\varepsilon>0\), it contains all hidden paths whose complete-data score lies
within \(\varepsilon\) of the optimum.

When comparing sequences of different lengths, it is often useful to work on a
normalized scale. Setting \(\varepsilon=T\eta\), one obtains
\[
\mathcal T^{\mathrm{norm}}_\eta(y_{1:T})
=
\left\{
s_{1:T}:
T^{-1}\Delta(s_{1:T};y_{1:T})\leq \eta
\right\}
=
\mathcal T_{T\eta}(y_{1:T}).
\]
The theoretical development uses \(\varepsilon\), while simulations and
applications may report both \(\varepsilon\) and \(\eta\). The normalized scale
\(\eta=\varepsilon/T\) should be interpreted as an average complete-data score
loss per time point. It is useful for reporting and for comparing trajectories
of similar lengths, but it is not a fixed posterior probability threshold
across different \(T\), because the relative posterior cutoff is
\(e^{-T\eta}\). For this reason, fixed-\(\eta\) summaries are used
descriptively, whereas entrance profiles and gap diagnostics provide the primary
scale-specific stability summaries.

\subsection{Projected state and transition tubes}

The full tube is a subset of \(\mathcal S^T\) and may contain many paths. For
interpretation, define its time-wise projection
\[
E_t(\varepsilon)
=
\left\{
k\in\mathcal S:
\exists s_{1:T}\in\mathcal T_\varepsilon(y_{1:T})
\text{ such that }s_t=k
\right\}.
\]
If \(E_t(\varepsilon)\) is a singleton, all paths in the tube agree on the
state at time \(t\). If \(|E_t(\varepsilon)|>1\), at least one near-optimal path
assigns a different state at that time.

For \(t=2,\ldots,T\), define the transition projection
\[
\mathcal A_t(\varepsilon)
=
\left\{
(i,j)\in\mathcal S^2:
\exists s_{1:T}\in\mathcal T_\varepsilon(y_{1:T})
\text{ such that }s_{t-1}=i,\ s_t=j
\right\}.
\]
The corresponding change-status projection is
\[
\mathcal C_t(\varepsilon)
=
\left\{
\mathbf 1\{i\neq j\}:
(i,j)\in\mathcal A_t(\varepsilon)
\right\}.
\]
Thus \(\mathcal C_t(\varepsilon)=\{0\}\) means all near-optimal paths remain in
the same state between \(t-1\) and \(t\), while
\(\mathcal C_t(\varepsilon)=\{1\}\) means all near-optimal paths change state.
The case \(\mathcal C_t(\varepsilon)=\{0,1\}\) indicates ambiguity in the
existence of a decoded change point.

The projections are exact projections of the pathwise tube, but they should not
be interpreted as generators of the full tube: a path using only
projected-admissible states or transitions need not itself belong to
\(\mathcal T_\varepsilon\). The Supplement gives explicit examples.

\subsection{Tube summaries}

The projected state width is
\[
w_t(\varepsilon)=|E_t(\varepsilon)|.
\]
A global mean width is
\[
W_\varepsilon
=
\frac{1}{T}\sum_{t=1}^{T}|E_t(\varepsilon)|.
\]
A state ambiguity proportion is
\[
A_\varepsilon^{\mathrm{state}}
=
\frac{1}{T}
\sum_{t=1}^{T}
\mathbf 1\{|E_t(\varepsilon)|>1\}.
\]
For \(K\geq 2\), define the normalized state concentration
\[
C_\varepsilon^{\mathrm{state}}
=
1-
\frac{1}{T(K-1)}
\sum_{t=1}^{T}
\left(|E_t(\varepsilon)|-1\right).
\]
For change-status stability, define
\[
R_\varepsilon^{\mathrm{cp}}
=
\frac{1}{T-1}
\sum_{t=2}^{T}
\mathbf 1\{|\mathcal C_t(\varepsilon)|=1\}.
\]
Large values of \(R_\varepsilon^{\mathrm{cp}}\) indicate that the tube agrees
on the presence or absence of decoded changes over most time points.

\subsection{Entrance tolerances and gap diagnostics}
\label{subsec:gap-diagnostics}

For interpretation across all values of \(\varepsilon\), it is useful to record
the tolerance at which each state or transition first enters a projected tube.
For a state \(k\) at time \(t\), define
\[
\tau_t^{\mathrm{state}}(k)
=
\psi^\star(y_{1:T})
-
\max_{s_{1:T}:s_t=k}
\psi(s_{1:T};y_{1:T}),
\]
and for a transition \(i\to j\) at time \(t\geq2\), define
\[
\tau_t^{\mathrm{trans}}(i,j)
=
\psi^\star(y_{1:T})
-
\max_{s_{1:T}:s_{t-1}=i,\ s_t=j}
\psi(s_{1:T};y_{1:T}).
\]
If the constrained set is empty, the corresponding tolerance is \(+\infty\).
Section~\ref{sec:computation} gives the exact dynamic programming computation
of these quantities.

Let \(\hat{s}_{1:T}\) be a selected Viterbi path, and define
\[
\hat B_t=\mathbf 1\{\hat{s}_t\neq \hat{s}_{t-1}\},
\qquad t=2,\ldots,T.
\]
The state alternative gap is
\[
g_t^{\mathrm{alt,state}}
=
\min_{k\neq \hat{s}_t}
\tau_t^{\mathrm{state}}(k),
\]
and the change-status gap is
\[
g_t^{\mathrm{cp}}
=
\min_{\substack{(i,j)\in\mathcal S^2:\\
\mathbf 1\{i\neq j\}\neq \hat B_t}}
\tau_t^{\mathrm{trans}}(i,j).
\]
Thus \(g_t^{\mathrm{alt,state}}\) is the minimum score loss needed to force a
state different from the selected Viterbi state at time \(t\), while
\(g_t^{\mathrm{cp}}\) is the minimum score loss needed to force a different
change-status from the selected Viterbi path. These gaps are often more
informative than a single fixed-width tube because they report the local
stability scale directly.

When multiple Viterbi paths exist, these gaps are defined relative to the
selected Viterbi path used for reporting. The tube itself is tie-invariant
because \(\mathcal T_0\) contains all Viterbi-optimal paths; only diagnostics
that compare to a selected path depend on the tie-breaking convention.

\section{Statistical Interpretation and Guarantees}
\label{sec:theory}

This section gives the statistical interpretation of the tropical Viterbi tube
and states the main guarantees used for inference. All probabilities are
conditional on the fitted HMM and on the observed sequence, unless otherwise
specified. Thus the results quantify decoding uncertainty under the fitted
model; they do not by themselves account for model misspecification or full
parameter uncertainty.

For compactness, write \(S=S_{1:T}\), \(Y=y\), and
\(\mathcal T_\varepsilon=\mathcal T_\varepsilon(y)\). Recall that
\[
\psi^\star(y)=\max_{s\in\mathcal S^T}\psi(s;y),
\qquad
\mathcal T_\varepsilon
=
\{s\in\mathcal S^T:\psi(s;y)\geq \psi^\star(y)-\varepsilon\}.
\]
The posterior distribution of the hidden path is
\[
P(S=s\mid Y=y)
=
\frac{\exp\{\psi(s;y)\}}
{\sum_{z\in\mathcal S^T}\exp\{\psi(z;y)\}}.
\]
The following results show that the tube is both a max-plus near-optimality set
and a posterior probability level set on the complete path space.

\subsection{Posterior superlevel interpretation}
\label{subsec:posterior-superlevel}

The first result gives the direct posterior interpretation of the score
tolerance \(\varepsilon\).

\begin{proposition}[Relative posterior superlevel set]
\label{prop:posterior-superlevel}
Under the fitted HMM,
\[
\mathcal T_\varepsilon
=
\left\{
s\in\mathcal S^T:
P(S=s\mid Y=y)
\geq
e^{-\varepsilon}
\max_{z\in\mathcal S^T}P(S=z\mid Y=y)
\right\}.
\]
Equivalently, if \(s^\star\) is any Viterbi-optimal path, then
\[
s\in\mathcal T_\varepsilon
\quad\Longleftrightarrow\quad
\frac{P(S=s\mid Y=y)}
     {P(S=s^\star\mid Y=y)}
\geq
e^{-\varepsilon}.
\]
\end{proposition}

Thus \(\varepsilon\) has a log posterior-odds interpretation. A path belongs to
\(\mathcal T_\varepsilon\) precisely when its posterior probability is at least
\(e^{-\varepsilon}\) times the posterior probability of a Viterbi path. In
particular, a fixed tolerance defines a relative posterior superlevel set. It
does not automatically define a credible set of prescribed posterior mass; that
requires calibration of \(\varepsilon\).

This distinction is important. The tube is not a collection of marginal
credible sets at individual times. It is a subset of the full path space
\(\mathcal S^T\). Its projections \(E_t(\varepsilon)\),
\(\mathcal A_t(\varepsilon)\), and \(\mathcal C_t(\varepsilon)\) summarize
which local features occur in at least one globally near-optimal trajectory.

\subsection{Posterior mass and likelihood contribution}
\label{subsec:tube-likelihood-mass}

Let
\[
\ell(y)
=
\log\sum_{s\in\mathcal S^T}\exp\{\psi(s;y)\}
\]
be the observed-data log-likelihood under the fitted model. Define the
tube-restricted log-likelihood contribution by
\[
\ell_\varepsilon^{\mathrm{tube}}(y)
=
\log\sum_{s\in\mathcal T_\varepsilon}
\exp\{\psi(s;y)\},
\]
and define the posterior mass of the tube by
\[
\Pi_\varepsilon^{\mathrm{tube}}(y)
=
P(S\in\mathcal T_\varepsilon\mid Y=y).
\]

\begin{theorem}[Tube likelihood mass]
\label{thm:tube-likelihood-mass}
For every \(\varepsilon\geq0\),
\[
\Pi_\varepsilon^{\mathrm{tube}}(y)
=
\exp\{
\ell_\varepsilon^{\mathrm{tube}}(y)-\ell(y)
\}.
\]
Consequently, the observed log-likelihood loss induced by restricting the
posterior path space to the tube is
\[
\Delta_\varepsilon^{\mathrm{obs}}(y)
=
\ell(y)-\ell_\varepsilon^{\mathrm{tube}}(y)
=
-\log \Pi_\varepsilon^{\mathrm{tube}}(y).
\]
Moreover, \(\Pi_\varepsilon^{\mathrm{tube}}(y)\) is nondecreasing in
\(\varepsilon\), while \(\Delta_\varepsilon^{\mathrm{obs}}(y)\) is
nonincreasing.
\end{theorem}

The quantity \(\Pi_\varepsilon^{\mathrm{tube}}\) measures how much posterior
path probability is carried by near-optimal complete trajectories. This is a
global measure of posterior concentration around the Viterbi optimum. A narrow
projected tube with large \(\Pi_\varepsilon^{\mathrm{tube}}\) indicates that
the posterior mass and the Viterbi optimum are concentrated in the same region
of path space. A narrow projected tube with small
\(\Pi_\varepsilon^{\mathrm{tube}}\) indicates that the Viterbi optimum may be
locally stable even though substantial posterior mass is carried by lower
scoring paths outside the tube.

\subsection{Posterior superlevel / HPD-threshold path regions}
\label{subsec:hpd-path-regions}

The previous proposition shows that every tube is a posterior superlevel set.
The next result states when such a tube becomes a credible region.

\begin{theorem}[Posterior superlevel / HPD-threshold path region]
\label{thm:hpd-path-set}
Let
\[
\varepsilon_\alpha^{\mathrm{HPD}}(y)
=
\inf
\left\{
\varepsilon\geq0:
\Pi_\varepsilon^{\mathrm{tube}}(y)\geq 1-\alpha
\right\}.
\]
Then
\[
\mathcal T_{\varepsilon_\alpha^{\mathrm{HPD}}}(y)
\]
is a posterior superlevel credible region, or HPD-threshold region, for the
complete hidden path, with posterior mass at least \(1-\alpha\). Because the
path space is discrete, the mass may be conservative, and boundary ties may
make minimal-mass HPD regions non-unique.
\end{theorem}

For an arbitrary tolerance \(\varepsilon\), the tube should be interpreted as a
relative posterior superlevel set. It becomes a posterior superlevel or
HPD-threshold credible region only after \(\varepsilon\) is chosen to attain a
target posterior mass. We use ``HPD'' in this threshold-set sense, not to claim
uniqueness or exact minimum mass under arbitrary tie-breaking. This distinction
is central to the use of tropical Viterbi tubes in applied work: small
tolerances describe local stability near the Viterbi optimum, whereas
mass-calibrated tolerances describe posterior credible regions on the full path
space.

\subsection{Projected tubes as simultaneous credible bands}
\label{subsec:projected-credible-bands}

Although the HPD tube is defined on \(\mathcal S^T\), its projections yield
time-indexed uncertainty bands for states, transitions, and change statuses.

\begin{theorem}[Projected credible bands]
\label{thm:projected-credible-bands}
If
\[
\Pi_\varepsilon^{\mathrm{tube}}(y)
=
P(S\in\mathcal T_\varepsilon\mid Y=y)
\geq 1-\alpha,
\]
then
\[
P\left(
S_t\in E_t(\varepsilon)
\text{ for all }t=1,\ldots,T
\mid Y=y
\right)
\geq 1-\alpha.
\]
Similarly,
\[
P\left(
(S_{t-1},S_t)\in\mathcal A_t(\varepsilon)
\text{ for all }t=2,\ldots,T
\mid Y=y
\right)
\geq 1-\alpha,
\]
and
\[
P\left(
\mathbf 1\{S_t\neq S_{t-1}\}\in\mathcal C_t(\varepsilon)
\text{ for all }t=2,\ldots,T
\mid Y=y
\right)
\geq 1-\alpha.
\]
\end{theorem}

Thus a posterior-mass-calibrated tube yields conservative simultaneous credible
bands for the projected state sequence, transition sequence, and change-status
sequence. These bands are conservative because projection loses pathwise
constraints: if \(S\in\mathcal T_\varepsilon\), then all its projected features
belong to the projected tube, but the reverse implication need not hold.
These projected bands are not claimed to be the smallest possible simultaneous
bands. They are conservative projections of a pathwise credible set.

The projected state bands should not be confused with pointwise marginal
credible sets. Posterior marginal probabilities
\[
p_t(k)=P(S_t=k\mid Y=y)
\]
describe local uncertainty at time \(t\). By contrast,
\(E_t(\varepsilon)\) contains states that occur at time \(t\) in at least one
complete path whose global score is near optimal. The resulting band is
therefore a projection of a pathwise object, not a collection of independent
local decisions.

\subsection{Properties of projected tube summaries}
\label{subsec:tube-summary-properties}

The projected summaries introduced in Section~\ref{sec:tropical-viterbi-tubes}
inherit the nested structure of the tube. It is useful to express this
multi-scale structure through entrance tolerances. Define
\[
\tau_t^{\mathrm{state}}(k)
=
\inf\{\varepsilon\geq0:k\in E_t(\varepsilon)\},
\]
and, for \(t=2,\ldots,T\),
\[
\tau_t^{\mathrm{trans}}(i,j)
=
\inf\{\varepsilon\geq0:(i,j)\in\mathcal A_t(\varepsilon)\}.
\]
Equivalently,
\[
\tau_t^{\mathrm{state}}(k)
=
\psi^\star(y)
-
\max_{\substack{s\in\mathcal S^T:\\ s_t=k}}
\psi(s;y),
\]
and
\[
\tau_t^{\mathrm{trans}}(i,j)
=
\psi^\star(y)
-
\max_{\substack{s\in\mathcal S^T:\\ s_{t-1}=i,\ s_t=j}}
\psi(s;y),
\]
with the convention that the maximum over an empty set is \(-\infty\), giving
entrance tolerance \(+\infty\). Section~\ref{sec:computation} shows how these
quantities are computed exactly by max-plus forward--backward recursions.

\begin{proposition}[Properties of projected tube summaries]
\label{prop:tube-summary-properties}
For every \(\varepsilon\geq0\),
\[
1\leq W_\varepsilon\leq K,
\qquad
0\leq A_\varepsilon^{\mathrm{state}}\leq 1,
\qquad
0\leq C_\varepsilon^{\mathrm{state}}\leq 1,
\qquad
0\leq R_\varepsilon^{\mathrm{cp}}\leq 1.
\]
If \(0\leq \varepsilon_1\leq \varepsilon_2\), then
\[
W_{\varepsilon_1}\leq W_{\varepsilon_2},
\qquad
A_{\varepsilon_1}^{\mathrm{state}}
\leq
A_{\varepsilon_2}^{\mathrm{state}},
\]
\[
C_{\varepsilon_1}^{\mathrm{state}}
\geq
C_{\varepsilon_2}^{\mathrm{state}},
\qquad
R_{\varepsilon_1}^{\mathrm{cp}}
\geq
R_{\varepsilon_2}^{\mathrm{cp}}.
\]
Moreover, as functions of \(\varepsilon\), these summaries are
right-continuous step functions whose jumps can occur only at state or
transition entrance tolerances.
\end{proposition}

The monotonicity is useful for interpretation. Increasing \(\varepsilon\)
relaxes the near-optimality requirement, so the projected state and transition
sets can only expand. Consequently, state width and ambiguity increase, while
state concentration and change-status stability decrease. The step-function
property implies that tube profiles are naturally multi-scale entrance
profiles. They change only when a new state or transition first becomes
compatible with a globally near-optimal path.

\subsection{Gap characterization of local stability}
\label{subsec:gap-characterization}

Let \(\hat{s}_{1:T}\) be a selected Viterbi path, and define its decoded
change-status sequence by
\[
\hat B_t
=
\mathbf 1\{\hat{s}_t\neq \hat{s}_{t-1}\},
\qquad t=2,\ldots,T.
\]
The state alternative gap at time \(t\) is
\[
g_t^{\mathrm{alt,state}}
=
\min_{k\neq \hat{s}_t}
\tau_t^{\mathrm{state}}(k),
\]
with the convention that the minimum over an empty set is \(+\infty\). The
change-status gap at time \(t\geq2\) is
\[
g_t^{\mathrm{cp}}
=
\min_{\substack{(i,j)\in\mathcal S^2:\\
\mathbf 1\{i\neq j\}\neq \hat B_t}}
\tau_t^{\mathrm{trans}}(i,j).
\]
These gaps measure the smallest loss in complete-data score required to force,
respectively, an alternative state at time \(t\) or an alternative change status
between times \(t-1\) and \(t\).

\begin{proposition}[Gap characterization of local stability]
\label{prop:gap-characterization}
For \(t=1,\ldots,T\),
\[
E_t(\varepsilon)=\{\hat{s}_t\}
\quad
\text{for every }0\leq\varepsilon<g_t^{\mathrm{alt,state}}.
\]
If \(g_t^{\mathrm{alt,state}}<\infty\), then
\[
|E_t(g_t^{\mathrm{alt,state}})|>1.
\]
Similarly, for \(t=2,\ldots,T\),
\[
\mathcal C_t(\varepsilon)=\{\hat B_t\}
\quad
\text{for every }0\leq\varepsilon<g_t^{\mathrm{cp}}.
\]
If \(g_t^{\mathrm{cp}}<\infty\), then
\[
|\mathcal C_t(g_t^{\mathrm{cp}})|>1.
\]
\end{proposition}

The boundary convention is important. Because the tube is defined using
\(\Delta(s;y)\leq\varepsilon\), an alternative state or change status enters
the corresponding projected set exactly at its entrance tolerance. A zero gap
therefore indicates an exact tie at the corresponding projected feature.

These gaps provide local robustness diagnostics. A large
\(g_t^{\mathrm{alt,state}}\) means that the decoded state at time \(t\) is
separated from all competing states by a large score gap. A large
\(g_t^{\mathrm{cp}}\) means that the presence or absence of a decoded change at
time \(t\) is stable against substantial score perturbations.

\subsection{Tube width and decoding instability}
\label{subsec:decoding-instability}

The tube also gives a deterministic stability bound. Let \(\psi\) and
\(\widetilde{\psi}\) be two complete-data score functions on the same path
space. Let \(\hat{s}\) be a Viterbi path for \(\psi\), and let
\(\widetilde{s}\) be a Viterbi path for \(\widetilde{\psi}\). Write
\(\mathcal T_\varepsilon^\psi\) for the tube constructed from the score
\(\psi\).

\begin{theorem}[Tube control of decoding instability]
\label{thm:decoding-instability}
If
\[
\sup_{s\in\mathcal S^T}
|\widetilde{\psi}(s)-\psi(s)|
\leq r,
\]
then
\[
\widetilde{s}\in\mathcal T_{2r}^{\psi}.
\]
Consequently, with all tube summaries computed from \(\psi\),
\[
\frac{1}{T}
\sum_{t=1}^{T}
\mathbf 1\{\widetilde{s}_t\neq \hat{s}_t\}
\leq
A_{2r}^{\mathrm{state}},
\]
and, writing
\[
\widetilde B_t
=
\mathbf 1\{\widetilde{s}_t\neq \widetilde{s}_{t-1}\},
\qquad
\hat B_t
=
\mathbf 1\{\hat{s}_t\neq \hat{s}_{t-1}\},
\]
\[
\frac{1}{T-1}
\sum_{t=2}^{T}
\mathbf 1\{\widetilde B_t\neq \hat B_t\}
\leq
1-R_{2r}^{\mathrm{cp}}.
\]
\end{theorem}

This theorem formalizes the connection between tube width and decoding
stability. If the fitted score is perturbed uniformly by at most \(r\), then
any Viterbi path under the perturbed score must lie inside the original tube at
radius \(2r\). Thus the projected tube at radius \(2r\) bounds how much the
decoded state sequence and decoded change-status sequence can change. In
applications, such perturbations may arise from numerical variation, local
changes in the fitted model, or approximate sensitivity analyses for estimated
parameters.

\subsection{Calibration of the tolerance}
\label{subsec:calibration}

The tolerance \(\varepsilon\) can be put on a statistical scale in several
ways. The three calibrations below answer different questions and should not be
treated as interchangeable.

\medskip\noindent\textit{Average set-valued calibration.}
Under the fitted model \(P_{\widehat\theta}\), one may simulate replicated
pairs \((S_{1:T},Y_{1:T})\) and target average time-wise coverage. Define
\[
\mathrm{Cov}_{\widehat\theta}^{\mathrm{state}}(\varepsilon)
=
E_{\widehat\theta}
\left[
\frac{1}{T}
\sum_{t=1}^{T}
\mathbf 1\{S_t\in E_t(\varepsilon;Y)\}
\right].
\]
An average state-coverage tolerance is
\[
\varepsilon_{\alpha}^{\mathrm{ave,state}}
=
\inf
\left\{
\varepsilon\geq0:
\mathrm{Cov}_{\widehat\theta}^{\mathrm{state}}(\varepsilon)
\geq 1-\alpha
\right\}.
\]
Similarly, with
\[
B_t(S)=\mathbf 1\{S_t\neq S_{t-1}\},
\]
one may define
\[
\mathrm{Cov}_{\widehat\theta}^{\mathrm{cp}}(\varepsilon)
=
E_{\widehat\theta}
\left[
\frac{1}{T-1}
\sum_{t=2}^{T}
\mathbf 1\{B_t(S)\in\mathcal C_t(\varepsilon;Y)\}
\right].
\]
Average calibration is useful for expected time-wise coverage and average set
size. It does not imply simultaneous coverage of the entire latent trajectory.

\medskip\noindent\textit{Simultaneous projected-band calibration.}
For a fixed observed sequence \(y\), define
\[
\mathrm{Sim}^{\mathrm{state}}_y(\varepsilon)
=
P\left(
S_t\in E_t(\varepsilon;y)
\text{ for all }t=1,\ldots,T
\mid Y=y
\right).
\]
The corresponding posterior simultaneous state-band tolerance is
\[
\varepsilon_{\alpha}^{\mathrm{sim,state}}(y)
=
\inf
\left\{
\varepsilon\geq0:
\mathrm{Sim}^{\mathrm{state}}_y(\varepsilon)\geq 1-\alpha
\right\}.
\]
Equivalently, define
\[
M_y^{\mathrm{state}}(S)
=
\max_{1\leq t\leq T}
\tau_t^{\mathrm{state}}(S_t;y).
\]
Then
\[
S_t\in E_t(\varepsilon;y)\text{ for all }t
\quad\Longleftrightarrow\quad
M_y^{\mathrm{state}}(S)\leq\varepsilon.
\]
Thus \(\varepsilon_{\alpha}^{\mathrm{sim,state}}(y)\) is a posterior quantile
of \(M_y^{\mathrm{state}}(S)\), up to discreteness.

For change statuses, define
\[
\tau_t^{\mathrm{cp}}(b;y)
=
\min_{\substack{(i,j)\in\mathcal S^2:\\
\mathbf 1\{i\neq j\}=b}}
\tau_t^{\mathrm{trans}}(i,j;y),
\qquad b\in\{0,1\},
\]
and
\[
M_y^{\mathrm{cp}}(S)
=
\max_{2\leq t\leq T}
\tau_t^{\mathrm{cp}}\{B_t(S);y\}.
\]
A simultaneous projected change-status tolerance can then be obtained from the
posterior quantile of \(M_y^{\mathrm{cp}}(S)\).

\medskip\noindent\textit{HPD path-mass calibration.}
HPD calibration targets the full path event
\[
S\in\mathcal T_\varepsilon(y).
\]
The tolerance is
\[
\varepsilon_\alpha^{\mathrm{HPD}}(y)
=
\inf
\left\{
\varepsilon\geq0:
\Pi_\varepsilon^{\mathrm{tube}}(y)\geq 1-\alpha
\right\}.
\]
Equivalently, if
\[
D_y(S)
=
\psi^\star(y)-\psi(S;y),
\qquad S\sim P(\cdot\mid Y=y),
\]
then
\[
\Pi_\varepsilon^{\mathrm{tube}}(y)
=
P\{D_y(S)\leq\varepsilon\mid Y=y\}.
\]
Thus \(\varepsilon_\alpha^{\mathrm{HPD}}(y)\) is the
\((1-\alpha)\)-quantile of the posterior path deficit, with the usual
conservatism due to discreteness.

The three calibrations have different interpretations. Average calibration
targets expected time-wise set coverage under a fitted data-generating model.
Simultaneous projected-band calibration targets posterior or model-based
coverage of all projected features at once. HPD path-mass calibration targets
posterior probability of the complete latent trajectory. All three are
conditional on the fitted HMM or on the fitted model used to generate
replicates. They are therefore model-based uncertainty summaries, not
distribution-free guarantees.

\begin{table}[t]
\centering
\caption{Calibration targets for tropical Viterbi tubes. The rows differ in
the target event, probability law, and computation used.}
\label{tab:calibration-targets}
\footnotesize
\setlength{\tabcolsep}{3pt}
\renewcommand{\arraystretch}{1.15}
\begin{tabular}{@{}L{0.15\linewidth}L{0.20\linewidth}L{0.20\linewidth}L{0.20\linewidth}L{0.19\linewidth}@{}}
\toprule
Component & Average set-valued & \shortstack[l]{Simultaneous\\projected state} & \shortstack[l]{Simultaneous\\projected\\change-status} & HPD path-mass\\
\midrule
Target event & \(T^{-1}\sum_t 1\{S_t\in E_t(\varepsilon;Y)\}\) & \(\{S_t\in E_t(\varepsilon;y)\ \forall t\}\) & \(\{B_t(S)\in\mathcal C_t(\varepsilon;y)\ \forall t\}\) & \(\{S\in\mathcal T_\varepsilon(y)\}\)\\
Probability law & Replicate law under fitted HMM or oracle simulation law & Posterior \(P(S\mid Y=y,\widehat\theta)\) or model-based replicate law & Posterior or model-based replicate law & Posterior \(P(S\mid Y=y,\widehat\theta)\)\\
Computation & Simulate \((S,Y)\), compute projected tubes & Quantile of \(M_y^{\mathrm{state}}(S)\) & Quantile of \(M_y^{\mathrm{cp}}(S)\) & Quantile of \(\Delta_y(S)=\psi^\star-\psi(S;y)\)\\
Exact? & MC & Exact if enumerated; MC with FFBS otherwise & Exact if enumerated; MC with FFBS otherwise & Not from projected recursion; MC in this paper\\
Used for & Average time-wise model-based coverage & Simultaneous projected bands & Simultaneous change-status bands & Full path credible region\\
\bottomrule
\end{tabular}
\renewcommand{\arraystretch}{1}
\end{table}

In the simulations below, average and simultaneous calibration are sometimes
evaluated under the known replicate-generating law to diagnose operating
characteristics. In a single applied data analysis, HPD and simultaneous
posterior calibrations require posterior path calculations, such as FFBS. These
calculations are distinct from the exact projected-tube recursion.

\section{Exact Computation and Implementation}
\label{sec:computation}

This section describes how to compute the projected tropical Viterbi tube
exactly. The main computational point is that the state and transition
projections of the tube do not require enumeration of the \(K^T\) hidden paths.
They can be obtained from one max-plus forward recursion and one max-plus
backward recursion. This gives all state and transition entrance tolerances, and
therefore all projected tubes over all values of \(\varepsilon\), in the same
asymptotic order as standard Viterbi decoding for dense transition matrices.
The exact \(O(TK^2)\) result in this section is for projected tube membership
and entrance tolerances. It is not an exact \(O(TK^2)\) algorithm for the full
posterior tube mass.

The section also separates this exact projected-tube computation from the
computation of the full posterior mass
\(\Pi_\varepsilon^{\mathrm{tube}}\). The latter is statistically useful, but it
requires summing posterior probability over complete paths satisfying a global
score constraint. It is not obtained by the same simple max-plus recursion.

\subsection{Max-plus forward and backward scores}
\label{subsec:maxplus-forward-backward}

Recall that the complete-data score of a hidden path \(s_{1:T}\) is
\[
\psi(s_{1:T};y_{1:T})
=
p_{s_1}+e_1(s_1)
+
\sum_{t=2}^{T}
\{a_t(s_{t-1},s_t)+e_t(s_t)\}.
\]
For \(t=1,\ldots,T\) and \(k\in\mathcal S\), define the max-plus forward score
\[
F_t(k)
=
\max_{s_{1:t-1}\in\mathcal S^{t-1}}
\psi(s_{1:t-1},s_t=k;y_{1:t}),
\]
Thus \(F_t(k)\) is the best partial score among all prefixes ending in state
\(k\) at time \(t\). It satisfies
\[
F_1(k)=p_k+e_1(k),
\]
and, for \(t=2,\ldots,T\),
\[
F_t(j)
=
e_t(j)
+
\max_{i\in\mathcal S}
\{F_{t-1}(i)+a_t(i,j)\}.
\]
The Viterbi score is then
\[
\psi^\star(y_{1:T})=\max_{k\in\mathcal S}F_T(k).
\]

Next define the max-plus backward score
\[
G_t(k)
=
\max_{s_{t+1:T}\in\mathcal S^{T-t}}
\sum_{u=t+1}^{T}
\{a_u(s_{u-1},s_u)+e_u(s_u)\},
\]
where the maximization is conditional on \(s_t=k\). Thus \(G_t(k)\) is the
best suffix contribution after time \(t\), given that the state at time \(t\)
is \(k\). The recursion is initialized by
\[
G_T(k)=0,
\]
and, for \(t=T-1,\ldots,1\),
\[
G_t(i)
=
\max_{j\in\mathcal S}
\{a_{t+1}(i,j)+e_{t+1}(j)+G_{t+1}(j)\}.
\]
The convention \(\log 0=-\infty\) is used throughout. Therefore impossible
states, emissions, or transitions are automatically excluded from the
maximizations.

\subsection{Exact projected tube theorem}
\label{subsec:exact-projected-tube}

The following theorem gives the computational characterization of projected
state and transition tubes.

\begin{theorem}[Exact projected tropical Viterbi tube]
\label{thm:exact-projected-tube}
Assume that at least one hidden path has finite complete-data score. For every
\(t=1,\ldots,T\), \(k\in\mathcal S\), and \(\varepsilon\geq0\),
\[
k\in E_t(\varepsilon)
\quad\Longleftrightarrow\quad
F_t(k)+G_t(k)\geq \psi^\star(y_{1:T})-\varepsilon.
\]
For every \(t=2,\ldots,T\), \(i,j\in\mathcal S\), and
\(\varepsilon\geq0\),
\[
(i,j)\in\mathcal A_t(\varepsilon)
\quad\Longleftrightarrow\quad
F_{t-1}(i)+a_t(i,j)+e_t(j)+G_t(j)
\geq
\psi^\star(y_{1:T})-\varepsilon.
\]
\end{theorem}

The theorem says that a state \(k\) belongs to the projected tube at time
\(t\) if and only if the best complete path constrained to visit \(k\) at
time \(t\) is within \(\varepsilon\) of the Viterbi score. Similarly, a
transition \((i,j)\) belongs to the projected transition tube at time \(t\) if
and only if the best complete path constrained to use that transition is within
\(\varepsilon\) of the optimum. The proof is given in the Supplement.

\subsection{Entrance tolerances}
\label{subsec:entrance-tolerances}

Theorem~\ref{thm:exact-projected-tube} motivates the state entrance tolerance
\[
\tau_t^{\mathrm{state}}(k)
=
\psi^\star(y_{1:T})-F_t(k)-G_t(k),
\qquad
t=1,\ldots,T,\quad k\in\mathcal S,
\]
and the transition entrance tolerance
\[
\tau_t^{\mathrm{trans}}(i,j)
=
\psi^\star(y_{1:T})
-
\{F_{t-1}(i)+a_t(i,j)+e_t(j)+G_t(j)\},
\]
for \(t=2,\ldots,T\) and \(i,j\in\mathcal S\). If the constrained state or
transition is impossible, the corresponding constrained score is \(-\infty\)
and the entrance tolerance is \(+\infty\).

The projected tubes can then be written as simple threshold sets:
\[
E_t(\varepsilon)
=
\{k\in\mathcal S:\tau_t^{\mathrm{state}}(k)\leq\varepsilon\},
\]
and
\[
\mathcal A_t(\varepsilon)
=
\{(i,j)\in\mathcal S^2:
\tau_t^{\mathrm{trans}}(i,j)\leq\varepsilon\}.
\]
The change-status projection follows by mapping transitions to indicators:
\[
\mathcal C_t(\varepsilon)
=
\{\mathbf 1\{i\neq j\}:(i,j)\in\mathcal A_t(\varepsilon)\}.
\]

Thus a single computation of the entrance tolerances gives the entire
multi-scale projected tube. Changing \(\varepsilon\) does not require rerunning
the dynamic program; it only requires thresholding the already computed
entrance profiles.

\subsection{Algorithm}
\label{subsec:algorithm}

Algorithm~\ref{alg:projected-tropical-tube} gives the exact computation of
\(\psi^\star\), the state entrance tolerances, and the transition entrance
tolerances.

\begin{algorithm}[H]
\caption{Exact projected tropical Viterbi tube}
\label{alg:projected-tropical-tube}
\begin{algorithmic}[1]
\Require Log-initial scores \(p_k\), log-transition scores \(a_t(i,j)\), log-emission scores \(e_t(k)\)
\Ensure Viterbi score \(\psi^\star\), state entrance tolerances \(\tau^{\mathrm{state}}\), transition entrance tolerances \(\tau^{\mathrm{trans}}\); not posterior tube mass

\Statex \textit{Forward max-plus recursion}
\For{\(k\in\mathcal S\)}
  \State \(F_1(k)\gets p_k+e_1(k)\)
\EndFor
\For{\(t=2,\ldots,T\)}
  \For{\(j\in\mathcal S\)}
    \State \(F_t(j)\gets e_t(j)+\max_{i\in\mathcal S}\{F_{t-1}(i)+a_t(i,j)\}\)
  \EndFor
\EndFor
\State \(\psi^\star\gets \max_{k\in\mathcal S}F_T(k)\)

\Statex \textit{Backward max-plus recursion}
\For{\(k\in\mathcal S\)}
  \State \(G_T(k)\gets 0\)
\EndFor
\For{\(t=T-1,\ldots,1\)}
  \For{\(i\in\mathcal S\)}
    \State \(G_t(i)\gets \max_{j\in\mathcal S}\{a_{t+1}(i,j)+e_{t+1}(j)+G_{t+1}(j)\}\)
  \EndFor
\EndFor

\Statex \textit{State entrance tolerances}
\For{\(t=1,\ldots,T\)}
  \For{\(k\in\mathcal S\)}
    \State \(\tau_t^{\mathrm{state}}(k)\gets \psi^\star-F_t(k)-G_t(k)\)
  \EndFor
\EndFor

\Statex \textit{Transition entrance tolerances}
\For{\(t=2,\ldots,T\)}
  \For{\(i\in\mathcal S\)}
    \For{\(j\in\mathcal S\)}
      \State \(\tau_t^{\mathrm{trans}}(i,j)\gets
      \psi^\star-\{F_{t-1}(i)+a_t(i,j)+e_t(j)+G_t(j)\}\)
    \EndFor
  \EndFor
\EndFor

\State \Return \(\psi^\star\), \(\tau^{\mathrm{state}}\), \(\tau^{\mathrm{trans}}\)
\end{algorithmic}
\end{algorithm}

For any tolerance \(\varepsilon\), the projected sets are recovered by
thresholding:
\[
E_t(\varepsilon)
=
\{k:\tau_t^{\mathrm{state}}(k)\leq\varepsilon\},
\]
\[
\mathcal A_t(\varepsilon)
=
\{(i,j):\tau_t^{\mathrm{trans}}(i,j)\leq\varepsilon\},
\]
and
\[
\mathcal C_t(\varepsilon)
=
\{\mathbf 1\{i\neq j\}:
\tau_t^{\mathrm{trans}}(i,j)\leq\varepsilon\}.
\]

\subsection{Multi-scale summaries}
\label{subsec:multiscale-summaries}

Given a grid of tolerances
\[
\mathcal G=\{\varepsilon_1,\ldots,\varepsilon_m\},
\]
all projected summaries can be computed by thresholding the entrance
tolerances. For example,
\[
w_t(\varepsilon)
=
|E_t(\varepsilon)|
=
\sum_{k=1}^{K}
\mathbf 1\{\tau_t^{\mathrm{state}}(k)\leq\varepsilon\}.
\]
Similarly, a transition width can be defined by
\[
v_t(\varepsilon)
=
|\mathcal A_t(\varepsilon)|
=
\sum_{i=1}^{K}\sum_{j=1}^{K}
\mathbf 1\{\tau_t^{\mathrm{trans}}(i,j)\leq\varepsilon\}.
\]
The global summaries introduced earlier are then computed as
\[
W_\varepsilon
=
\frac{1}{T}\sum_{t=1}^{T}w_t(\varepsilon),
\]
\[
A_\varepsilon^{\mathrm{state}}
=
\frac{1}{T}\sum_{t=1}^{T}
\mathbf 1\{w_t(\varepsilon)>1\},
\]
\[
C_\varepsilon^{\mathrm{state}}
=
1-
\frac{1}{T(K-1)}
\sum_{t=1}^{T}\{w_t(\varepsilon)-1\},
\]
and
\[
R_\varepsilon^{\mathrm{cp}}
=
\frac{1}{T-1}
\sum_{t=2}^{T}
\mathbf 1\{|\mathcal C_t(\varepsilon)|=1\}.
\]
If transition ambiguity is of interest, one may also report
\[
V_\varepsilon
=
\frac{1}{T-1}
\sum_{t=2}^{T}
|\mathcal A_t(\varepsilon)|.
\]

A useful default grid \(\mathcal G\) is given by empirical quantiles of the
finite entrance tolerances
\[
\{\tau_t^{\mathrm{state}}(k)\}
\cup
\{\tau_t^{\mathrm{trans}}(i,j)\}.
\]
Another natural choice is to use all finite entrance tolerances themselves.
This produces the exact step-function profiles of the projected tube
summaries, at the cost of a potentially larger grid.

\subsection{Observed likelihood and posterior marginal quantities}
\label{subsec:ordinary-forward}

The quantities in this subsection use ordinary posterior recursions. The
log-likelihood of the observations
\[
\ell(y_{1:T})
=
\log
\sum_{s\in\mathcal S^T}
\exp\{\psi(s;y_{1:T})\}
\]
is computed by the usual log-sum-exp forward recursion. Let
\[
L_1(k)=p_k+e_1(k),
\]
and, for \(t=2,\ldots,T\),
\[
L_t(j)
=
e_t(j)
+
\operatorname{logsumexp}_{i\in\mathcal S}
\{L_{t-1}(i)+a_t(i,j)\}.
\]
Then
\[
\ell(y_{1:T})
=
\operatorname{logsumexp}_{k\in\mathcal S} L_T(k).
\]
Posterior marginal probabilities \(P(S_t=k\mid Y=y)\), local entropy, and
posterior path sampling can be obtained by standard forward-backward or
forward-filtering backward-sampling calculations. These quantities are useful
for comparison with the tropical Viterbi tube, but they are not needed to
compute the projected tube itself.

\subsection{Posterior mass of the full tube is a separate pathwise computation}
\label{subsec:posterior-tube-mass-computation}

The exact projected tube is computed by Theorem~\ref{thm:exact-projected-tube}.
The full posterior mass
\[
\Pi_\varepsilon^{\mathrm{tube}}(y)
=
P(S\in\mathcal T_\varepsilon(y)\mid Y=y)
\]
is a different computational object. It can be written as
\[
\Pi_\varepsilon^{\mathrm{tube}}(y)
=
\frac{
\sum_{s\in\mathcal T_\varepsilon(y)}
\exp\{\psi(s;y)\}
}{
\sum_{z\in\mathcal S^T}
\exp\{\psi(z;y)\}
},
\]
but the numerator sums over complete paths satisfying the global constraint
\[
\psi(s;y)\geq \psi^\star(y)-\varepsilon.
\]
This global score constraint does not factor locally in the same way as the
projected membership conditions in Theorem~\ref{thm:exact-projected-tube}.
Therefore \(\Pi_\varepsilon^{\mathrm{tube}}\) is not obtained by the same
\(O(TK^2)\) max-plus forward-backward recursion.

Several strategies are available, depending on \(K\), \(T\), and the required
accuracy.

\medskip\noindent\textit{Exact enumeration.}
For small \(K\) and \(T\), one may enumerate all \(K^T\) paths, compute their
scores, and sum the posterior mass of those with deficit at most
\(\varepsilon\). This is useful for validation, but it is not scalable.

\medskip\noindent\textit{\(M\)-best path approximation.}
If a list of the \(M\) highest-scoring paths is available, one may compute the
partial mass
\[
\underline{\Pi}_{\varepsilon,M}^{\mathrm{tube}}
=
\frac{
\sum_{m=1}^{M}
\mathbf 1\{s^{(m)}\in\mathcal T_\varepsilon\}
\exp\{\psi(s^{(m)};y)\}
}{
\exp\{\ell(y)\}
}.
\]
This is a lower bound on \(\Pi_\varepsilon^{\mathrm{tube}}\). It is most useful
when posterior mass is concentrated in a moderate number of high-scoring paths.

\medskip\noindent\textit{Posterior path sampling.}
If \(S^{(1)},\ldots,S^{(B)}\) are draws from the posterior distribution
\(P(S\mid Y=y)\), then
\[
\widehat{\Pi}_{\varepsilon}^{\mathrm{tube}}
=
\frac{1}{B}
\sum_{b=1}^{B}
\mathbf 1
\left\{
\psi(S^{(b)};y)\geq \psi^\star(y)-\varepsilon
\right\}
\]
is an unbiased Monte Carlo estimator of
\(\Pi_\varepsilon^{\mathrm{tube}}\), conditional on exact posterior sampling.
Monte Carlo uncertainty should be reported when this estimator is used for HPD
calibration or for likelihood-loss summaries.

\medskip\noindent\textit{Sequential or discretized score methods.}
For longer sequences, one may approximate the tube mass using sequential Monte
Carlo or by augmenting a dynamic program with a discretized score-deficit
coordinate. These approaches trade computational cost and approximation error
against the ability to estimate path-mass quantities beyond the projected tube.

The main exact computational contribution of this paper is the projected tube
and its entrance profiles. Whenever the full posterior tube mass is approximated,
the approximation method and its numerical uncertainty should be reported
separately from statistical Monte Carlo uncertainty in simulation or calibration
experiments.
Thus, throughout the empirical sections, projected tubes, entrance tolerances
and gap profiles are exact outputs of the max-plus recursion, whereas posterior
tube masses and HPD tolerances are estimated posterior path quantities whenever
FFBS is used.

\subsection{Complexity and storage}
\label{subsec:complexity}

For a dense \(K\times K\) transition matrix, the forward max-plus recursion
costs \(O(TK^2)\), and the backward max-plus recursion also costs \(O(TK^2)\).
Computing all state entrance tolerances costs \(O(TK)\), and computing all
transition entrance tolerances costs \(O(TK^2)\). Therefore the exact projected
state and transition tubes, together with all entrance profiles, are computed in
\[
O(TK^2)
\]
time for dense transitions.

The memory cost is \(O(TK)\) for storing the forward and backward score arrays.
State entrance tolerances require \(O(TK)\) storage. If all transition entrance
tolerances are stored, they require \(O(TK^2)\) storage. If only state tubes or
change-status summaries are required, transition tolerances can be processed
one time point at a time to reduce memory usage.
For large \(K\), applications that only require change-status summaries can
compute transition tolerances one time point at a time and discard the full
\(K\times K\) array after updating \(\mathcal C_t(\varepsilon)\).

For sparse transition graphs, let \(m_t\) be the number of allowed transitions
from time \(t-1\) to time \(t\). Then the forward and backward recursions can be
implemented over the allowed edges, giving time complexity
\[
O\left(\sum_{t=2}^{T}m_t\right),
\]
up to the \(O(TK)\) emission terms. The transition entrance tolerances are then
computed only for allowed transitions; impossible transitions have entrance
tolerance \(+\infty\).

\subsection{Numerical conventions}
\label{subsec:numerical-conventions}

All computations should be performed on the log scale. Impossible events are
represented by \(-\infty\), and maximizations over empty feasible sets return
\(-\infty\). Entrance tolerances are nonnegative in exact arithmetic because
\(\psi^\star\) is the maximum complete-data score. In floating-point arithmetic,
very small negative entrance tolerances may occur because of numerical roundoff;
these can be truncated to zero.

For numerical thresholding, it is useful to define
\[
E_t^{\mathrm{num}}(\varepsilon)
=
\{k:\tau_t^{\mathrm{state}}(k)\leq \varepsilon+\xi\},
\]
and similarly for transitions, where \(\xi\) is a small numerical tolerance.
The choice of \(\xi\) should be reported if it materially affects the projected
sets. In typical double-precision implementations, \(\xi\) can be chosen close
to machine precision on the scale of the accumulated log-scores.

\FloatBarrier

\section{Simulation Study}
\label{sec:simulation}
\suppressfloats[t]

Using the computation above, we report a compact simulation study designed to
check the algorithmic implementation and to show how projected tubes behave
across stable, ambiguous, near-tie, and misspecified decoding regimes. The
results are model-based for the HMM used in each experiment. Monte Carlo
standard errors are computed across simulation replicates.

The projected tube algorithm was first checked by brute-force enumeration. For
\((K,T)=(2,6)\) and \((K,T)=(3,6)\), all \(K^T\) paths were enumerated, and
the state and transition projections agreed with the max-plus algorithm for
all tested tolerances. The main oracle simulation then used \(T=100\) and 200
replicates in six scenarios: well-separated emissions, overlapping emissions,
localized transitions, a harder localized-transition setting, a near-tie
setting, and a misspecified Student-\(t\) emission setting. We used the fine
normalized tolerance grid
\[
\begin{aligned}
\eta \in \{&
0,0.0001,0.00025,0.0005,0.001,0.0015,0.002,0.0025,\\
&0.003,0.004,0.005,0.0075,0.01,0.015,0.02,0.03,0.05\},
\end{aligned}
\]
with \(\varepsilon=\eta T\). Posterior path calculations for representative
trajectories used FFBS with 50,000 or 100,000 posterior samples, depending on
the scenario. Full simulation tables and auxiliary figures are reported in the
Supplement. The complete data-generating parameters for all scenarios are given
in Supplement Table S0, including
transition matrices, emission parameters, initial distributions, and the fitted
model used in the misspecified Student-\(t\) scenario.

Table~\ref{tab:accuracy-entropy-v4} summarizes point decoding accuracy and
posterior entropy. The localized-transition and well-separated scenarios are
stable, with mean state accuracies 0.997 and 0.987 and normalized entropies
0.011 and 0.037. The overlapping and near-tie scenarios are more ambiguous,
with mean state accuracies 0.810 and 0.753 and normalized entropies 0.387 and
0.617. The misspecified Student-\(t\) scenario is the most difficult, with mean
state accuracy 0.617 and mean normalized entropy 0.685.

\begin{table}[!t]
\centering
\caption{Point decoding accuracy and posterior entropy for the main
simulation. Values are Monte Carlo means with replicate-level standard errors
in parentheses.}
\label{tab:accuracy-entropy-v4}
\begin{tabular}{@{}lcccc@{}}
\toprule
Scenario & $K$ & State accuracy & Change accuracy & Normalized entropy\\
\midrule
Well separated & 3 & 0.987 (0.0009) & 0.982 (0.0013) & 0.037 (0.0011)\\
Overlapping & 3 & 0.810 (0.0074) & 0.921 (0.0021) & 0.387 (0.0052)\\
Localized transitions & 2 & 0.997 (0.0004) & 0.995 (0.0008) & 0.011 (0.0006)\\
Localized transitions, hard & 2 & 0.975 (0.0018) & 0.979 (0.0013) & 0.092 (0.0032)\\
Near tie & 2 & 0.753 (0.0128) & 0.967 (0.0013) & 0.617 (0.0089)\\
Misspecified Student-t & 3 & 0.617 (0.0130) & 0.897 (0.0023) & 0.685 (0.0050)\\
\bottomrule
\end{tabular}
\end{table}

\begin{figure}[t]
\centering
\includegraphics[width=\linewidth]{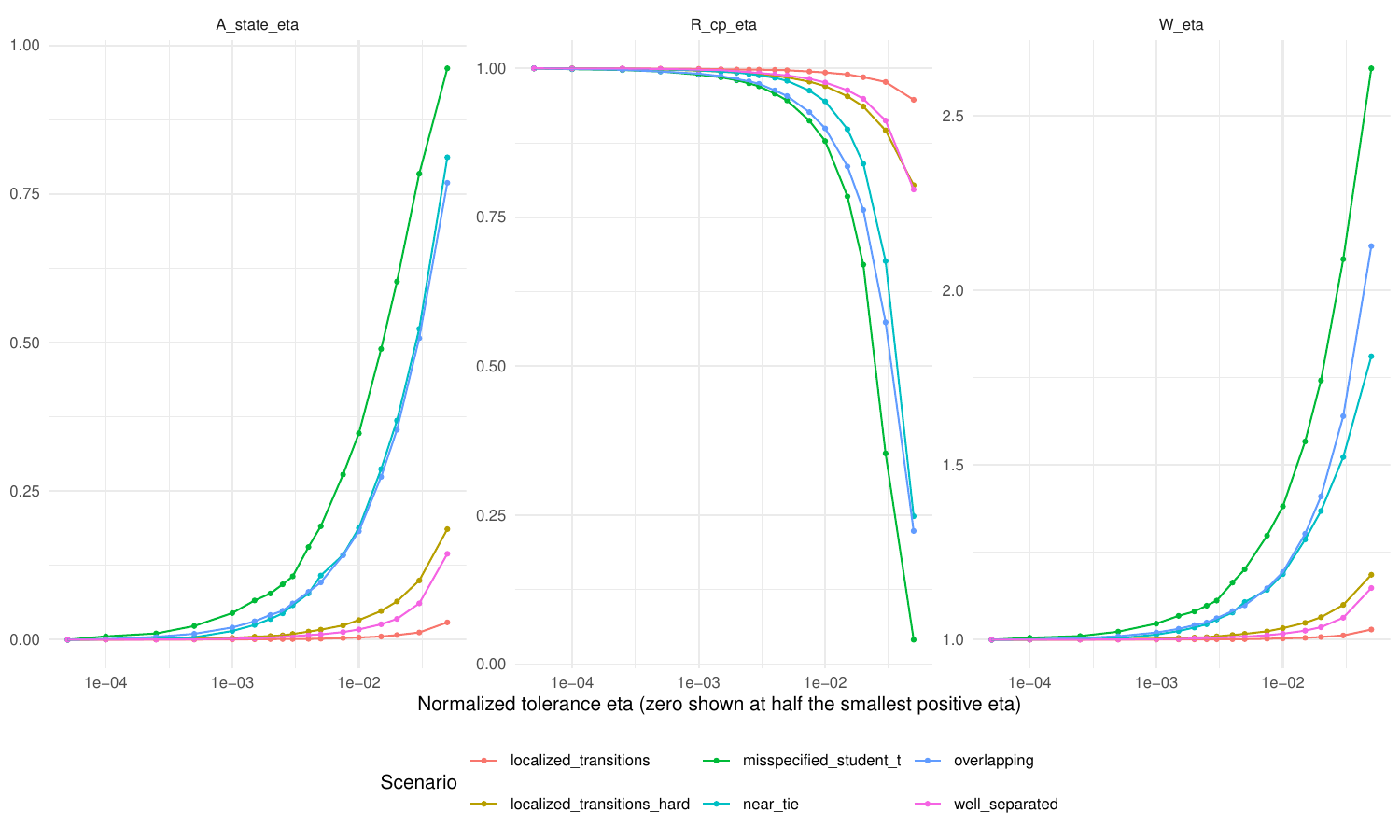}
\caption{Multi-scale projected tube profiles over the fine normalized
tolerance grid. The zero-tolerance value is included as the Viterbi-optimal
baseline; positive tolerances show how projected state and change-status
ambiguity enter as the tube expands. Fixed \(\eta\) values are descriptive;
entrance profiles show the tolerance scale at which ambiguity enters.}
\label{fig:multiscale-profiles-v4}
\end{figure}

The primary set-valued diagnostic is the multi-scale entrance profile rather
than a single fixed tolerance. Figure~\ref{fig:multiscale-profiles-v4} shows
the projected state width \(W_\eta\), state ambiguity
\(A_\eta^{\mathrm{state}}\), and change-status robustness
\(R_\eta^{\mathrm{cp}}\) over the fine grid. At \(\eta=0.005\), the stable
scenarios have nearly singleton projections: \(W_\eta=1.002\) for localized
transitions and \(W_\eta=1.009\) for well-separated emissions. The projected
tubes expand earlier for overlapping emissions, near-ties, and misspecification,
with \(W_\eta=1.098\), 1.108, and 1.202, respectively. Thus the entrance
profile records the scale at which decoding ambiguity enters.

Tube width can saturate, especially in ambiguous or misspecified settings. Gap
diagnostics instead measure the local score loss required to force a competing
state or a competing change-status. For the decoded state \(\hat s_t\),
\[
g_t^{\mathrm{alt,state}}
= \min_{k\ne \hat s_t}\tau_t^{\mathrm{state}}(k),
\]
and \(g_t^{\mathrm{cp}}\) is defined analogously for the opposite
change-status. Larger values indicate more stable local decoding.
Table~\ref{tab:gap-summary-v4} shows that these gaps separate the regimes
clearly. The mean state alternative gap is about 15.07 for localized
transitions, 8.75 for well-separated emissions, 8.21 for the hard
localized-transition scenario, 3.21 for overlapping emissions, 3.03 for the
near-tie scenario, and 1.87 for the misspecified Student-\(t\) scenario.

\begin{table}[t]
\centering
\caption{Gap diagnostics in the main simulation. Values are Monte Carlo means
with replicate-level standard errors in parentheses.}
\label{tab:gap-summary-v4}
\scriptsize
\setlength{\tabcolsep}{2.5pt}
\resizebox{\linewidth}{!}{
\begin{tabular}{@{}lcccccc@{}}
\toprule
Diagnostic & \shortstack{Well\\sep.} & Overlap. & \shortstack{Local\\trans.} & \shortstack{Local\\hard} & \shortstack{Near\\tie} & \shortstack{Student-\\\(t\)}\\
\midrule
Mean \(g^{\mathrm{alt,state}}\) & 8.753 (0.0476) & 3.205 (0.0582) & 15.068 (0.0348) & 8.212 (0.0561) & 3.033 (0.0896) & 1.874 (0.0629)\\
Median \(g^{\mathrm{alt,state}}\) & 8.914 (0.0457) & 2.963 (0.0762) & 15.394 (0.0390) & 8.850 (0.0617) & 2.804 (0.1081) & 1.699 (0.0723)\\
Mean \(g^{\mathrm{cp}}\) & 7.504 (0.0433) & 3.449 (0.0431) & 12.806 (0.0389) & 7.759 (0.0494) & 3.772 (0.0656) & 2.580 (0.0421)\\
Median \(g^{\mathrm{cp}}\) & 7.658 (0.0450) & 3.454 (0.0511) & 13.279 (0.0462) & 8.539 (0.0531) & 3.831 (0.0715) & 2.636 (0.0448)\\
Mean \(g^{\mathrm{alt,state}}/T\) & 0.0875 (0.00048) & 0.0320 (0.00058) & 0.1507 (0.00035) & 0.0821 (0.00056) & 0.0303 (0.00090) & 0.0187 (0.00063)\\
Mean \(g^{\mathrm{cp}}/T\) & 0.0750 (0.00043) & 0.0345 (0.00043) & 0.1281 (0.00039) & 0.0776 (0.00049) & 0.0377 (0.00066) & 0.0258 (0.00042)\\
\bottomrule
\end{tabular}
}
\begin{flushleft}
\footnotesize
Column labels abbreviate well separated, overlapping, localized transitions,
localized transitions hard, near tie, and misspecified Student-\(t\).
\end{flushleft}
\end{table}

Average, simultaneous projected, and HPD path-mass calibration answer different
questions. Average time-wise model-based calibration targets
\(T^{-1}\sum_t 1\{S_t\in E_t(\varepsilon;Y)\}\) under the fitted or oracle
replicate law. At \(\eta=0.005\), the well-separated scenario has average state
coverage 0.991 but simultaneous state coverage 0.465. The localized-transition
scenario has average state coverage 0.998 and simultaneous state coverage
0.845. In contrast, overlapping emissions have average state coverage 0.853
and simultaneous state coverage zero at this tolerance. Thus high average
time-wise model-based coverage can leave simultaneous projected-band coverage
low.

In the oracle simulation, simultaneous projected calibration is evaluated using
the replicate-level maximum of the entrance tolerance of the true latent state
sequence. In a single observed sequence, the analogous posterior simultaneous
tolerance is obtained from posterior draws of \(S\mid Y=y\), using
\(M_y^{\mathrm{state}}(S)=\max_t\tau_t^{\mathrm{state}}(S_t;y)\). This target
therefore produces wider projected bands. At nominal 0.95, simultaneous state
coverage is 0.975 in the well-separated scenario, with average state size
1.241, and 0.935 in the overlapping scenario, with average state size 2.540.

HPD path-mass calibration is pathwise: for a fixed observed sequence and HMM,
FFBS posterior paths were sampled and their Viterbi deficits
\[
\Delta(S)=\psi^\star-\psi(S;y)
\]
were used to estimate the tolerance whose tube carries a target posterior mass.
For localized transitions, the posterior mass at \(\varepsilon=0\) is already
0.988. For overlapping emissions, 0.95 HPD calibration requires
\(\eta=0.2218\) and saturates the projected state, transition, and
change-status sizes at 3.000, 9.000, and 2.000. The Monte Carlo standard errors
reported for HPD mass estimates quantify uncertainty in the estimated mass at
the selected tolerance; they do not fully quantify quantile-selection
uncertainty for \(\eta_\alpha^{\mathrm{HPD}}\).

Finally, the posterior tube mass
\[
\Pi_\varepsilon^{\mathrm{tube}}
=P(S\in \mathcal T_\varepsilon\mid Y=y)
\]
summarizes the complete path mass carried by the tube and is distinct from
local projected width. At \(\eta=0.05\), the estimated posterior tube mass is
0.990 for localized transitions, 0.708 for well-separated emissions, 0.446 for
near-ties, 0.160 for overlapping emissions, and 0.00537 for the misspecified
Student-\(t\) setting. Thus projected tube summaries and complete path mass
answer different questions.

The fitted-versus-oracle, entropy-gap, length-sensitivity, and perturbation
experiments are reported in the Supplement. They show that fitted-model
decoding is close to oracle decoding in stable regimes but more sensitive in
overlapping and near-tie regimes. They also show that fixed normalized
tolerances can widen with sequence length, and that the perturbation
experiment confirms the deterministic containment theorem with conservative
bounds relative to observed Hamming instability. Table~\ref{tab:fitted-oracle-v4}
summarizes the fitted-versus-oracle comparison at \(\eta=0.005\). This length
effect is not a failure of the method: it reflects the fact that \(\eta\) is an
average score-loss scale, while the posterior relative threshold is
\(e^{-T\eta}\). Therefore fixed-\(\eta\) displays are descriptive, and gap
profiles are the preferred summaries when comparing sequences of different
lengths.

\begin{table}[t]
\centering
\caption{Fitted-versus-oracle comparison at \(T=500\) and \(\eta=0.005\).
Values are Monte Carlo means over 200 fitted-model replicates. The last column
summarizes the qualitative sensitivity of the fitted projected tube relative to
the oracle tube.}
\label{tab:fitted-oracle-v4}
\resizebox{\linewidth}{!}{
\begin{tabular}{lrrrl}
\toprule
Scenario & Viterbi disagreement & Change-status disagreement & Mean \(W_\eta\) fitted-oracle diff. & Comment\\
\midrule
Well separated & 0.004 & 0.004 & -0.003 & stable\\
Localized transitions, hard & 0.021 & 0.012 & 0.001 & mostly stable\\
Overlapping & 0.151 & 0.034 & -0.070 & more sensitive\\
Near tie & 0.300 & 0.083 & 0.093 & most sensitive\\
\bottomrule
\end{tabular}
}
\end{table}

We next apply the same projected summaries to a public bat movement data set.

\FloatBarrier

\section{Applied Illustration: Decoding Uncertainty in Animal Movement HMMs}
\label{sec:application}

\subsection{Data and ecological question}

We illustrate the method using the public Movebank Data Repository package of
\citet{Hurme2019MovebankData}, associated with the Movement Ecology study of
\citet{Hurme2019MovementEcology}. The data contain GPS trajectories and
on-board acoustic recordings from Mexican fish-eating bats
(\emph{Myotis vivesi}) at Isla Partida Norte, Mexico. The applied question is
not whether an HMM can produce a foraging/commuting segmentation. It can. The
question is which parts of that segmentation are stable as complete latent
paths, and whether those stability summaries agree with an external ecological
signal. We therefore evaluate the tube not by asking whether it outperforms
Viterbi at detecting feeding buzzes, but by asking whether it identifies which
parts of the Viterbi segmentation are pathwise stable or ambiguous.

The external signal is the occurrence of feeding buzzes in the acoustic record.
We use feeding buzzes as independent evidence of prey-capture attempts, not as a
continuous annotation of foraging. In particular, the absence of a feeding buzz
does not imply absence of foraging. This distinction is important: the
comparison is informative about enrichment, but it does not provide sensitivity
or specificity for the latent behavioural states.

\subsection{Preprocessing and fitted movement HMM}

The application scripts join GPS records to deployment metadata, project
longitude and latitude to UTM zone 12N, remove locations close to the island
using the 250 m reference filter described in the original study, and discard
trips with fewer than 100 retained locations. The resulting analysis uses
\BatNumTrips{} trips, \BatNumLocations{} retained GPS locations,
\BatNumSteps{} retained step observations, and \BatNumBuzzes{} feeding buzzes
occurring at \BatNumBuzzLocations{} locations. The mean retained trip
duration is \BatMeanDurationMin{} minutes.

We fit a two-state HMM to step length and turning angle. Step length has a
state-dependent gamma distribution, and turning angle has a state-dependent von
Mises distribution with mean direction fixed at zero. This is a standard
movement-HMM specification used in animal movement software such as
\texttt{momentuHMM} \citep{McClintockMichelot2018}. In this application the
likelihood is evaluated directly, so that the fitted log-initial vector,
homogeneous log-transition matrix, and log-emission matrix are available
without package-specific extraction rules. The HMM is used as a working
movement model for conditional decoding uncertainty. We do not claim that this
two-state HMM is the unique or best ecological model for these trajectories. It
is used as a standard, interpretable movement-HMM working model to study
conditional decoding uncertainty. The tube summaries below are therefore
conditional on this fitted model.

The fitted log-likelihood is \BatHMMLogLik{} and the AIC is \BatHMMAIC{}. State
labels are assigned after fitting: the foraging state is the state with shorter
steps and less directional turning, while commuting is the state with longer
steps and turning angles concentrated near zero. The fitted foraging state has
mean step length \BatForagingStepMean{} m and von Mises concentration
\BatForagingKappa{}, while the commuting state has mean step length
\BatCommutingStepMean{} m and concentration \BatCommutingKappa{}.

\subsection{Standard decoding summaries}

The Viterbi path gives one globally optimal behavioural segmentation for each
trip. It is therefore a natural baseline for ecological interpretation. We also
compute posterior state marginals by the forward--backward recursion and the
normalized posterior entropy at each retained step observation. These summaries
answer different questions. The Viterbi path is a single coherent latent
history, posterior marginals give pointwise state probabilities, and entropy
summarizes local dispersion in those pointwise probabilities.

For comparison with the tube, we form two standard threshold summaries:
Viterbi foraging versus Viterbi commuting, and marginal categories
\(P(S_t=\mathrm{foraging}\mid Y)\geq 0.9\), \(0.1<P(S_t=\mathrm{foraging}\mid
Y)<0.9\), and \(P(S_t=\mathrm{foraging}\mid Y)\leq 0.1\). Entropy is grouped
using the fixed thresholds 0.1 and 0.5. These thresholds are useful descriptive
benchmarks, but they remain local summaries. They do not ask whether an entire
alternative path remains close to the Viterbi optimum.

\subsection{Tropical Viterbi tube summaries}

For each trip we compute max-plus forward scores, max-plus backward scores,
state entrance tolerances, transition entrance tolerances, projected state tube
widths, change-status tubes, state alternative gaps, and change-status gaps.
The normalized tolerance grid is
\[
0,\ 0.0005,\ 0.001,\ 0.0025,\ 0.005,\ 0.01,\ 0.02,\ 0.05,\ 0.10,
\]
with \(\varepsilon=\eta T\) for each trip. At the reference tolerance
\(\eta=\BatEtaRef\), the projected state tube partitions the retained step
observations into \BatTubeRobustForagingPctEta\% robust foraging,
\BatTubeAmbiguousPctEta\% ambiguous, and
\BatTubeRobustCommutingPctEta\% robust commuting. These are not additional
state labels fitted from the acoustic data. They are projections of the
near-optimal complete paths under the fitted HMM.
For the seven retained trips, \(\eta=0.005\) corresponds to
\(\varepsilon=T\eta\) values ranging from 1.42 to 6.96 complete-score units.
The corresponding relative posterior cutoff \(e^{-\varepsilon}\) ranges from
approximately 0.24 to 0.001, illustrating why \(\eta\) should be interpreted as
an average score-loss scale rather than a fixed posterior probability
threshold. This value is used as a descriptive reference because it yields a
nontrivial but not saturated ambiguous category: 17.9\% of step observations
are ambiguous at \(\eta=0.005\), compared with 54.6\% at \(\eta=0.01\) and
79.2\% at \(\eta=0.02\). The multi-scale profiles, rather than this single
value, are the primary sensitivity summary.

\begin{figure}[t]
\centering
\includegraphics[width=\linewidth]{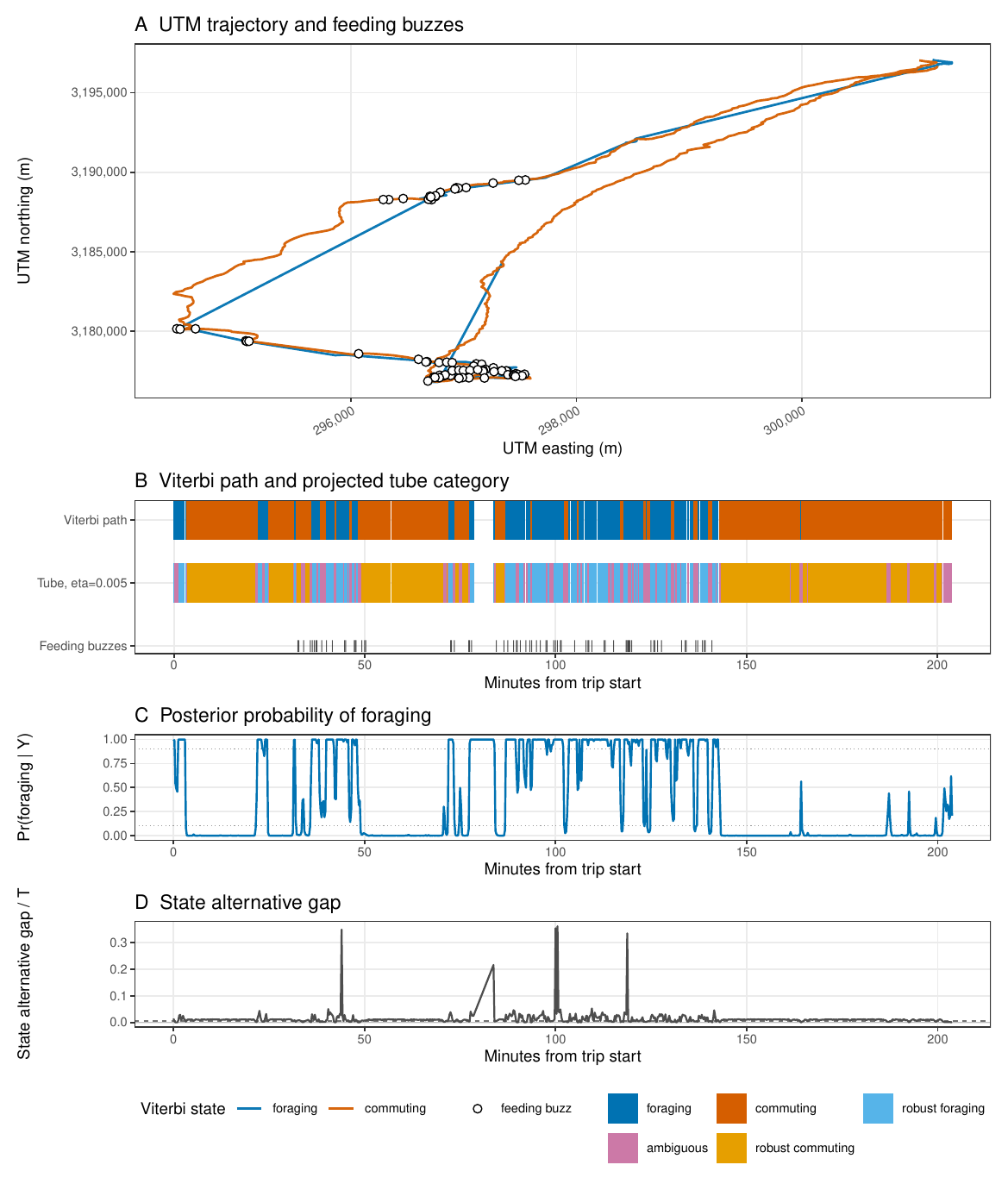}
\caption{Representative bat trip. Panel A shows the UTM trajectory coloured by
the Viterbi state, with locations containing feeding buzzes marked as open
points. Panel B
compares the Viterbi segmentation with the projected tube category at
\(\eta=\BatEtaRef\). Panel C shows the posterior probability of foraging.
Panel D shows the normalized state alternative gap, the score scale at which
an alternative state enters the near-optimal path set. Feeding buzzes provide
external evidence of prey-capture attempts, not continuous ground truth for the
latent state. Tube categories are projections at \(\eta=\BatEtaRef\); they are
not fitted from acoustic buzzes.}
\label{fig:bat-main-application}
\end{figure}

Figure~\ref{fig:bat-main-application} shows the representative trip
\BatRepresentativeTrip. The large map displays the Viterbi segmentation and
locations with feeding buzzes. The timeline then contrasts the binary Viterbi
ribbon with the three-way tube ribbon at \(\eta=\BatEtaRef\). The posterior
curve shows local state probability, while the gap panel shows the score scale
at which an alternative state enters the near-optimal set.

\FloatBarrier

The fitted HMM parameters used to compute these summaries are reported in
Table~\ref{tab:bat-hmm}.

\begin{table}[t]
\centering
\caption{Fitted two-state movement HMM for the bat application. Step length is
modelled by a gamma distribution and turning angle by a von Mises distribution
with mean direction zero. Stationary probabilities and expected dwell times are
computed from the fitted transition matrix. The last two columns are fitted
transition probabilities from the row state to the indicated destination state.}
\label{tab:bat-hmm}
\resizebox{\linewidth}{!}{
\begin{tabular}{lrrrrrrr}
\toprule
State & Step mean & Step SD & $\kappa$ & Stationary prob. & Dwell & To foraging & To commuting\\
\midrule
foraging & 51.8 & 49.0 & 0.001 & 0.350 & 12.1 & 0.918 & 0.082\\
commuting & 84.8 & 20.9 & 9.720 & 0.650 & 22.6 & 0.044 & 0.956\\
\bottomrule
\end{tabular}
}
\end{table}

\FloatBarrier

\subsection{External acoustic comparison}

The acoustic data provide an external check on the interpretation of the
movement-only HMM. At \(\eta=\BatEtaRef\), robust foraging step observations
have buzz rate \BatBuzzRateRobustForaging{} and enrichment
\BatBuzzEnrichmentRobustForaging{} relative to the pooled buzz rate, with 95\%
trip-level bootstrap interval \BatBuzzEnrichmentRobustForagingCI. Ambiguous
step observations have buzz rate \BatBuzzRateAmbiguous{} and enrichment
\BatBuzzEnrichmentAmbiguous{} with interval \BatBuzzEnrichmentAmbiguousCI.
Robust commuting step observations have buzz rate \BatBuzzRateRobustCommuting{}
and enrichment \BatBuzzEnrichmentRobustCommuting{} with interval
\BatBuzzEnrichmentRobustCommutingCI.

The bootstrap intervals resample the seven retained trips with replacement. For
each bootstrap sample, step observations and buzz counts are aggregated within
tube categories, and enrichment is recomputed relative to the pooled buzz rate
in the resampled trips. Categories absent in a bootstrap resample are treated as
missing for that category-specific interval.

We also fit an aggregated rate model with trip fixed effects and an offset for
the number of retained step observations in each trip-category cell. The
response is the number of buzzes in each trip-category cell, the offset is the
log number of retained step observations in that cell, and trip indicators are
included as fixed effects. Relative to robust foraging, the fitted
\BatRateModelType{} rate ratio is
\BatRateRatioAmbiguousVsRobustForaging{} for ambiguous cells
\BatRateRatioAmbiguousVsRobustForagingCI{} and
\BatRateRatioCommutingVsRobustForaging{} for robust commuting cells
\BatRateRatioCommutingVsRobustForagingCI. This model is descriptive and is not
used as the primary uncertainty analysis because there are only seven trips;
the bootstrap intervals are the primary uncertainty summary.

\begin{table}[t]
\centering
\caption{Feeding buzz summaries by projected tube category at
\(\eta=\BatEtaRef\). Buzz rate is the number of feeding buzzes per
retained step observation. Enrichment is relative to the pooled buzz rate over
all retained step observations. Intervals are trip-level percentile bootstrap
intervals obtained by resampling retained trips with replacement.}
\label{tab:bat-buzz-validation}
\begin{tabular}{lrrrr}
\toprule
Tube category & Step obs. & Buzzes & Buzz rate (95\% CI) & Enrichment (95\% CI)\\
\midrule
robust foraging & 1,487 & 171 & 0.115 (0.075, 0.159) & 2.25 (1.73, 2.85)\\
ambiguous & 940 & 59 & 0.063 (0.027, 0.114) & 1.23 (0.84, 1.56)\\
robust commuting & 2,837 & 39 & 0.014 (0.008, 0.025) & 0.27 (0.16, 0.44)\\
\bottomrule
\end{tabular}
\end{table}

\begin{figure}[t]
\centering
\includegraphics[width=0.86\linewidth]{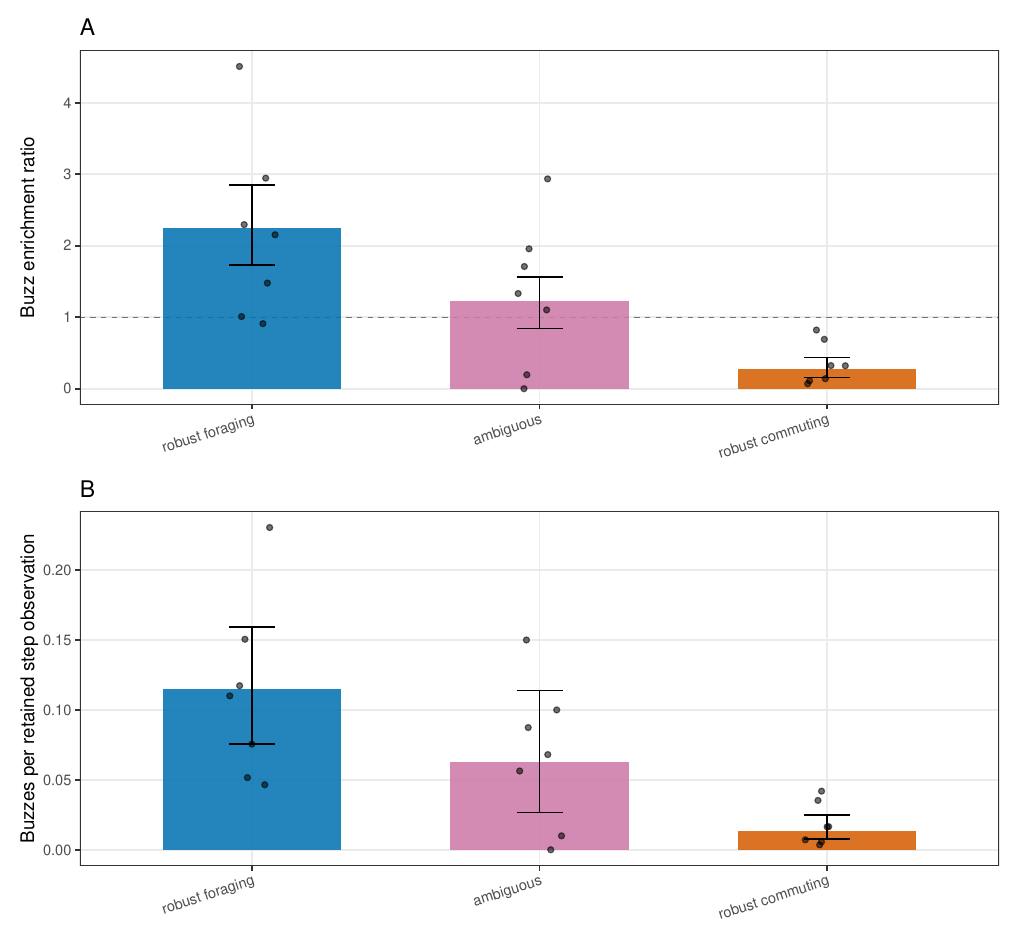}
\caption{External acoustic comparison by projected tube category at
\(\eta=\BatEtaRef\). Panel A shows enrichment relative to the pooled buzz
rate; Panel B shows the buzz rate per retained step observation. Points are
trip-level category summaries and intervals are trip-level percentile bootstrap
intervals. Feeding buzzes are independent evidence of prey-capture attempts.}
\label{fig:bat-buzz-validation}
\end{figure}

\FloatBarrier

\subsection{Comparison with Viterbi and posterior marginal summaries}

The enrichment pattern is not a statement that only the tube can find foraging
segments. The Viterbi and marginal summaries also separate higher and lower
feeding buzz rates: Viterbi foraging has enrichment
\BatViterbiForagingEnrichment{}, Viterbi commuting has enrichment
\BatViterbiCommutingEnrichment{}, high marginal foraging has enrichment
\BatMarginalForagingEnrichment{}, and high marginal commuting has enrichment
\BatMarginalCommutingEnrichment. The point of the tube is different. It asks
which parts of those decoded periods remain stable when the object of
comparison is a complete latent path.

This distinction is visible in the cross-classification of Viterbi state and
tube category. Within Viterbi-foraging observations, robust foraging and
ambiguous tube segments have nearly identical buzz rates
\BatViterbiForagingRobustRate{} and \BatViterbiForagingAmbiguousRate,
respectively. Within Viterbi-commuting observations, however, ambiguous tube
segments have buzz rate \BatViterbiCommutingAmbiguousRate{}, compared with
\BatViterbiCommutingRobustRate{} for robust commuting segments. Thus, for these
data, the tube chiefly refines the commuting part of the Viterbi path by
identifying intervals where a near-optimal alternative path can enter and where
the acoustic signal is less depleted than in robust commuting periods.

\begin{table}[t]
\centering
\caption{Cross-classification of Viterbi state and projected tube category at
\(\eta=\BatEtaRef\). Buzz rate is the number of feeding buzzes per retained
step observation, and enrichment is relative to the pooled buzz rate over all
retained step observations.}
\label{tab:bat-viterbi-tube-cross}
\begin{tabular}{llrrrr}
\toprule
Viterbi state & Tube category & Step obs. & Buzzes & Buzz rate & Enrichment\\
\midrule
foraging & robust foraging & 1,487 & 171 & 0.115 & 2.25\\
foraging & ambiguous & 336 & 39 & 0.116 & 2.27\\
foraging & robust commuting & 0 & 0 & -- & --\\
commuting & robust foraging & 0 & 0 & -- & --\\
commuting & ambiguous & 604 & 20 & 0.033 & 0.65\\
commuting & robust commuting & 2,837 & 39 & 0.014 & 0.27\\
\bottomrule
\end{tabular}
\end{table}

Entropy supplies another useful comparison. Low entropy does not imply robust
foraging; it may also correspond to highly certain commuting. Conversely, high
entropy identifies locally mixed posterior probabilities but does not encode
whether a globally coherent alternative path is available.
Table~\ref{tab:bat-standard-comparison} reports the corresponding buzz
summaries for Viterbi, tube, marginal, and entropy categories.

\begin{table}[t]
\centering
\caption{Feeding buzz summaries for standard decoding summaries and projected
tube categories. Buzz rate is the number of feeding buzzes per retained
step observation. These summaries are not mutually exclusive and answer
different inferential questions. The tube categories summarize projected
membership in globally near-optimal complete paths, whereas marginal and
entropy categories are local posterior summaries.}
\label{tab:bat-standard-comparison}
\resizebox{\linewidth}{!}{
\begin{tabular}{llrrr}
\toprule
Summary & Category & Step obs. & Buzz rate & Enrichment\\
\midrule
Viterbi state & Viterbi commuting & 3,441 & 0.017 & 0.34\\
Viterbi state & Viterbi foraging & 1,823 & 0.115 & 2.25\\
Tropical tube category & ambiguous & 940 & 0.063 & 1.23\\
Tropical tube category & robust commuting & 2,837 & 0.014 & 0.27\\
Tropical tube category & robust foraging & 1,487 & 0.115 & 2.25\\
Posterior marginal category & high marginal commuting & 3,159 & 0.014 & 0.27\\
Posterior marginal category & high marginal foraging & 1,623 & 0.119 & 2.33\\
Posterior marginal category & marginal ambiguous & 482 & 0.068 & 1.34\\
Posterior entropy category & high entropy & 458 & 0.072 & 1.41\\
Posterior entropy category & low entropy & 4,308 & 0.045 & 0.89\\
Posterior entropy category & moderate entropy & 498 & 0.080 & 1.57\\
\bottomrule
\end{tabular}
}
\end{table}

\FloatBarrier

\subsection{Sensitivity to the tolerance}

The value \(\eta=\BatEtaRef\) is a reference tolerance, not a universal cutoff.
Tube summaries should be read together with their multi-scale profiles. As
\(\eta\) increases from 0.0025 to 0.02, the ambiguous category increases from
\BatEtaLowAmbiguousPct\% to \BatEtaHighAmbiguousPct\% of retained step
observations. Over the same range, robust foraging enrichment changes from
\BatEtaLowRobustForagingEnrichment{} to
\BatEtaHighRobustForagingEnrichment{}, while robust commuting enrichment
changes from \BatEtaLowRobustCommutingEnrichment{} to
\BatEtaHighRobustCommutingEnrichment. The robust commuting category is small at
the largest tolerance, so that endpoint should be interpreted cautiously.

\begin{figure}[t]
\centering
\includegraphics[width=0.82\linewidth]{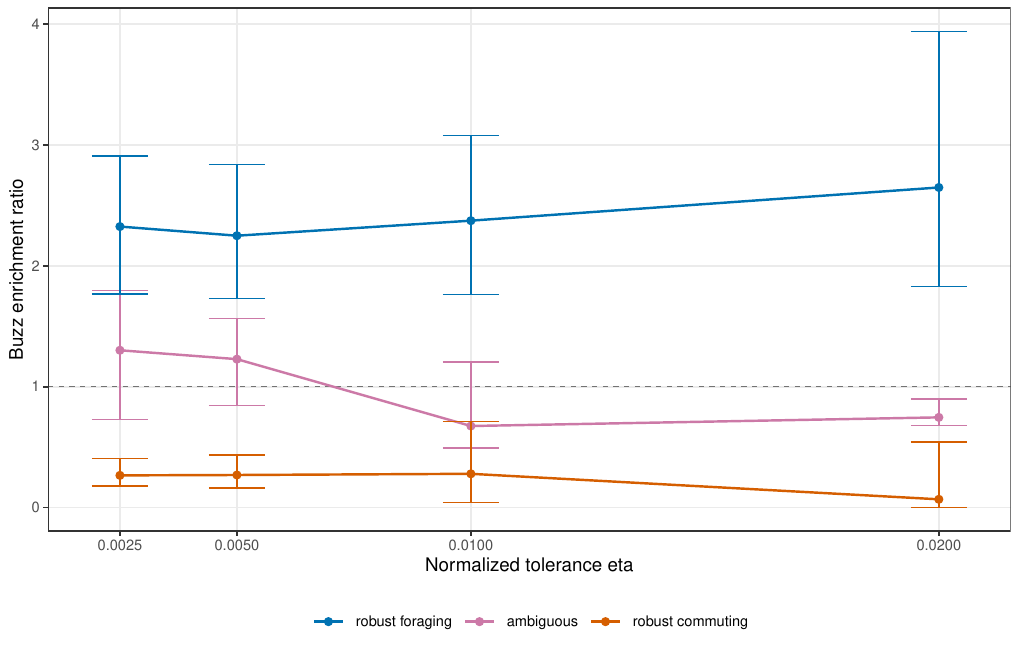}
\caption{Sensitivity of feeding buzz enrichment to the normalized tube
tolerance. Points show enrichment estimates by tube category, and intervals are
trip-level percentile bootstrap intervals. The horizontal line marks no
enrichment. The tolerance controls the size of the near-optimal path set and
should be interpreted as a scale parameter. The ambiguous category increases
from 8.6\% to 79.2\% over this grid; robust commuting is small at
\(\eta=0.02\), so the endpoint should be interpreted cautiously.}
\label{fig:bat-eta-sensitivity}
\end{figure}

\FloatBarrier

\subsection{Applied interpretation}

The application supports a cautious interpretation. The tropical Viterbi tube
does not replace the Viterbi path, the posterior marginals, or the entropy. It
adds a pathwise stability layer to them. In this fitted working HMM, robust foraging
tube segments are enriched for feeding buzzes, robust commuting segments are
depleted, and ambiguous segments identify portions of the decoded history where
a complete alternative path remains close to the Viterbi score. The main
applied value is therefore not a new behavioural label, but a partition of the
decoded trajectory into stable foraging, ambiguous, and stable commuting
segments that can be checked against independent acoustic evidence.

Taken together, these summaries show that, under the fitted movement HMM, the
pathwise tube separates robust and ambiguous parts of the decoding in a way that
is consistent with observed prey-capture attempts.

\FloatBarrier

\section{Discussion}
\label{sec:discussion}

The main contribution is an exact projected stability analysis of Viterbi
decoding: for a fitted HMM, the method identifies which local states,
transitions and change statuses are compatible with globally near-optimal
complete paths. The tropical Viterbi tube does not replace Viterbi decoding. It
adds a pathwise stability layer around the Viterbi path by retaining the same
complete-data score and asking which complete trajectories remain nearly
optimal.

This distinguishes the tube from posterior marginals and entropy. Marginals
and entropy describe local posterior uncertainty, while the tropical Viterbi
tube identifies local alternatives that are compatible with at least one
globally coherent near-optimal path. The contribution is therefore a pathwise
summary of decoding uncertainty, rather than another pointwise classification.

The tube is also related to \(k\)-best and list Viterbi methods in that all
concern high-scoring complete paths. However, the aim here is not to enumerate
alternatives. It is to compute exact projections and entrance tolerances of the
entire score-threshold set. Enumeration can approximate the full tube but is not
required for projected membership.

The bat movement application illustrates this added layer. Robust foraging tube
segments are enriched in feeding buzzes, robust commuting tube segments are
depleted, and ambiguous segments are useful especially within parts of the
Viterbi commuting path. Thus the applied contribution is not a new behavioural
classifier, but a refinement of the decoded trajectory into stable and
ambiguous portions that can be compared with external evidence of
prey-capture attempts.

The tolerance \(\varepsilon\) is interpretable on a log-posterior-odds scale.
Small values describe paths nearly tied with the Viterbi optimum, while larger
values include lower-scoring but still coherent trajectories. Fixed normalized
values \(\eta=\varepsilon/T\) can be useful for reporting, but entrance
tolerances and gap diagnostics are often more informative because they show the
scale at which specific states or change statuses become ambiguous.

The simulation results also show that average time-wise projected coverage,
simultaneous projected coverage, and HPD path mass answer distinct questions.
Average calibration is useful for local summaries, simultaneous calibration is
appropriate when the full projected band must contain the trajectory at all
times, and HPD calibration targets posterior mass in complete-path space.
Consequently, a pathwise HPD tube can be statistically well calibrated even
when its local projections become broad or saturated. This should be interpreted
as evidence of genuine path uncertainty under the fitted model, not as a
failure of the projection algorithm, and it does not contradict fixed-\(\eta\)
stability summaries that focus on smaller neighbourhoods of the Viterbi path.

The main inferential limitation is that all posterior statements are
conditional on the fitted HMM. They do not automatically incorporate parameter
uncertainty or protect against model misspecification. Perturbation results and
fitted-versus-oracle simulations can diagnose some sensitivity, but they do not
replace a full treatment of uncertainty in the model parameters. In the bat
application, feeding buzzes are also imperfect external evidence: they indicate
observed prey-capture attempts, but their absence does not prove the absence of
foraging.

A second limitation is computational. Projected tubes and their entrance
tolerances are exact in \(O(TK^2)\) for dense transition matrices, but the full
posterior mass \(\Pi_\varepsilon^{\mathrm{tube}}\) is not computed by the same
simple max-plus recursion. HPD path-mass calibration therefore requires
enumeration, sampling, approximation, or augmented dynamic programming. In the
simulation study, these pathwise quantities were approximated by FFBS, and the
reported posterior tube masses and HPD thresholds therefore include Monte Carlo
error from posterior path sampling.

The projections \(E_t(\varepsilon)\), \(\mathcal A_t(\varepsilon)\), and
\(\mathcal C_t(\varepsilon)\) are exact projections of the complete path tube,
but the projections do not reconstruct the full tube. They are designed as
interpretable summaries of pathwise uncertainty. Similarly, the deterministic
perturbation bounds give containment guarantees and can be conservative; they
should not be read as sharp predictions of empirical Hamming instability.
Therefore a projected band should be read as a conservative summary of pathwise
uncertainty, not as a Cartesian-product representation of admissible paths.

Several extensions are natural. The same pathwise idea could be developed for
hidden semi-Markov models, HMM step-selection functions, covariate-dependent
transition models, and other structured latent-state time series. Incorporating
parameter uncertainty is a natural extension, but it is not required for the
conditional decoding question addressed here. Bootstrap or Bayesian parameter
draws could be combined with the projected tube by recomputing entrance profiles
across fitted parameter values.

\section*{Data and code availability}
The bat movement data are publicly available from the Movebank Data Repository
\citep{Hurme2019MovebankData}. This preprint package includes the manuscript
source, figure files, supplementary material, and the reproducibility archive
prepared for the AOAS submission. The archive contains the R scripts for the
simulations and bat application, processed-data construction scripts, figure and
table generation scripts, seeds, package information, runtime notes, and
instructions for obtaining the public Movebank data.

\begin{supplement}
\stitle{Supplementary material for ``Tropical Viterbi Tubes for Decoding Uncertainty in Hidden Markov Models''}
\sdescription{The supplement contains proofs of the main results, additional geometric interpretations, algorithmic validation details, simulation diagnostics, posterior tube mass approximation details, and additional details for the bat movement application.}
\end{supplement}

\end{document}